\documentclass[a4paper,fleqn,usenatbib]{mnras}

\usepackage{newtxtext,newtxmath}

\usepackage[T1]{fontenc}
\usepackage{ae,aecompl}

\usepackage{graphicx}	
\usepackage{amsmath}	

\usepackage{enumitem}
\setlist[itemize]{leftmargin=*}

\usepackage{epsfig}
\usepackage{multirow}
\usepackage{float}
\usepackage{footmisc}

\def\apjl{ApJL}
\def\apjs{ApJS}
\def\aap{A\&A}

\def\xmm{{\it XMM-Newton}}
\def\asca{{\it ASCA}}

\def\rosat{{\it ROSAT}}
\def\nustar{{\it NuSTAR}}
\def\swift{{\it Swift}}
\def\rx04{RX J0439.6-5311}
\def\rxj0134{RX J0134.2-4258}
\def\rej1034{RE J1034+396}
\def\1h07{1H 0707-495}
\def\pg12{PG 1244+026}
\def\grs1915{GRS 1915+105}
\def\phl1811{PHL 1811}

\title[\rxj0134\ -- I. Peculiar X-ray Spectra \& Variability ]{Multi-wavelength Campaign on the Super-Eddington NLS1 \rxj0134\ -- I. Peculiar X-ray Spectra and Variability}

\author[C. Jin, et al.]{Chichuan Jin$^{1,2}$\thanks{E-mail: ccjin@nao.cas.cn},
Chris Done$^{3}$,
Martin Ward$^{3}$,
Francesca Panessa$^{4}$,
Bo Liu$^{1}$,
Heyang Liu$^{1}$
\smallskip
\\
$^{1}$National Astronomical Observatories, Chinese Academy of Sciences, 20A Datun Road, Beijing 100101, China\\
$^{2}$School of Astronomy and Space Sciences, University of Chinese Academy of Sciences, 19A Yuquan Road, Beijing 100049, China\\
$^{3}$Centre for Extragalactic Astronomy, Department of Physics, University of Durham, South Road, Durham DH1 3LE, UK\\
$^{4}$INAF - Istituto di Astrofisica e Planetologia Spaziali (IAPS-INAF), Via del Fosso del Cavaliere 100, I-00133 Roma, Italy\\
}

\date{prepared for MNRAS}

\pubyear{2021}

\begin{document}
\label{firstpage}
\pagerange{\pageref{firstpage}--\pageref{lastpage}}
\maketitle

\begin{abstract}
We have conducted a new long-term multi-wavelength campaign on 
one of the most super-Eddington narrow-line Seyfert 1s (NLS1s) known, namely \rxj0134. In this first paper, we report deep simultaneous X-ray observations performed by \xmm\ and \nustar\ in 2019-12-19, during which \rxj0134\ was fortuitously at one of its lowest X-ray flux states. However, there is a clear rise above 4~keV which implies that the intrinsic source flux may be higher. The X-ray spectra observed between 1996 and 2019 show drastic variability, probably due to complex, variable absorption along the line of sight.
Unusually, the soft X-ray excess appears extremely weak in all these spectra, even when the hard X-ray spectrum has a steep slope of $\Gamma \simeq 2.2$. We explore the spectral-timing properties of the new (low X-ray flux) and archival (high X-ray flux) \xmm\ data, fitting their time-average, rms and lag spectra simultaneously. The variability spectra indicate the presence of a very weak soft X-ray Comptonisation component, whose shape is similar to the soft excess in normal super-Eddington NLS1s, but with flux relative to the power law which is lower by more than one order of magnitude. 
Above 4~keV the low-flux data are dominated by a different component, 
which lags with respect to the lower energy emission. This is consistent with an origin of reflection or partial covering absorption from low ionisation material located within 100 $R_{\rm g}$. We interpret this as further indication of the presence of a clumpy disc wind.
\end{abstract}




\section{Introduction}
\label{sec-intro}
\subsection{Narrow-line Seyfert 1 Galaxies}
The emission of active galactic nuclei (AGN) originates from accretion onto a super-massive black hole (SMBH) in the center of their host galaxies. The fundamental parameters determining the AGN emission include the mass, mass accretion rate and spin of the SMBH, though inclination angle also plays a key role in the observed properties of AGN (\citealt{Antonucci.1993}). Narrow-line Seyfert 1 (NLS1) galaxies are a sub-class of AGN, 
characterized by an H$\beta$ line-width which is narrow compared to standard broad line AGN, but still broader than the narrow lines. The narrow lines such as 
[O {\sc iii}]$\lambda$5007 line are weak, while the FeII emission is strong(\citealt{Osterbrock.1985}; \citealt{Boroson.2002}). Comparing with the entire AGN population, NLS1s have smaller black hole masses of $10^{6-7}~M_{\odot}$ and higher Eddington ratios (e.g. \citealt{Pounds.1995}; \citealt{Mathur.2001}; \citealt{Jin.2012a}), although some NLS1s may accrete with normal/low Eddington ratios (e.g. NGC 4051: \citealt{McHardy.2004}).

The typical X-ray spectrum of NLS1s comprises a steep 2-10 keV power law and a strong soft X-ray excess below 2 keV (e.g. \citealt{Boller.1996}; \citealt{Brandt.1997}). \citet{Gallo.2006} divided NLS1s into two sub-types, namely the X-ray {\it complex} NLS1s and X-ray {\it simple} NLS1s. The X-ray spectral shape of the {\it complex} NLS1s show strong features at Fe~K$\alpha$, and significant variability, while {\it simple} NLS1s tend to have a much smoother X-ray spectrum with less variability. The soft excess of {\it complex} NLS1s can be modelled by an ionized disc reflection component (e.g. \citealt{Miniutti.2004}; \citealt{Ross.2005}; \citealt{Crummy.2006}; \citealt{Fabian.2013}), or by the absorption and scattering of disc wind material (e.g. \citealt{Miller.2007}; \citealt{Turner.2007}; \citealt{Tatum.2012}; \citealt{Hagino.2016}).
The soft excess of X-ray {\it simple} NLS1s can be best modelled by a low-temperature Comptonisation component (e.g. \citealt{Laor.1997}; \citealt{Magdziarz.1998}; \citealt{Done.2012}; \citealt{Jin.2013, Jin.2016, Jin.2017a, Jin.2021}).

The X-ray {\it simple} and {\it complex} NLS1s both have similar low black hole masses ($10^{6-7} M_\odot$) and 
high Eddington ratios ($L_{\rm bol}/L_{\rm Edd}\gtrsim 1$, where $L_{\rm bol}$ and $L_{\rm Edd}$ are the bolometric luminosity and Eddington luminosity). Thus they should have similar intrinsic accretion disc properties. But 
super-Eddington accretion is likely to lead to a puffed up inner disc (funnel geometry), and to power a strong wind (\citealt{Ohsuga.2011, Takeuchi.2014, Jiang.2016}). Inclination effects can then be important. 
The wind, especially when it is clumpy, can introduce significant time dependent viewing-angle effects, thereby providing a natural explanation for the apparent differences between the X-ray {\it complex} NLS1s (i.e. higher inclination angles, lines of sight penetrate through the wind) and the X-ray {\it simple} NLS1s (i.e. lower inclination angles, clean lines of sight) (\citealt{Hagino.2016, Jin.2017b}).

\subsection{The Enigmatic NLS1 Galaxy: \rxj0134}
\rxj0134\ was first discovered by the {\it ROSAT} all sky survey (\citealt{Voges.1999}). It is classified as a NLS1 galaxy at a redshift of $z=0.237$. It has a typical optical spectrum of an extreme NLS1s, including extremely weak [O {\sc iii}] $\lambda$5007 lines, narrow Balmer lines (H$\beta$ FWHM = 1160 km s$^{-1}$), strong Fe {\sc ii} and a steeply rising optical/UV continuum (\citealt{Grupe.2000}). The black hole mass was reported in \citet{Grupe.2004}, where the equations in \citet{Kaspi.2000} were applied to derive the virial mass of  $M_{\rm BH}=1.5\times10^{7}M_{\odot}$. The bolometric luminosity was derived from the broadband spectral energy distribution of \rxj0134\ as reported in \citet{Grupe.2010}, and so the Eddington ratio was estimated to be $L_{\rm bol}/L_{\rm Edd}=10.0$. This Eddington ratio is the second highest in the \swift\ AGN sample of \citet{Grupe.2010}, only slightly lower than \rx04, which is another extremely super-Eddington NLS1, whose multi-wavelength properties have been studied in detail by \citet{Jin.2017a} and \citet{Jin.2017b}. Since the black hole mass and Eddington ratio are both crucial parameters for \rxj0134, we have used independent methods to verify and constrain them better, as we will present in Section~\ref{sec-short-xvar} and the next paper (Jin et al. in preparation, hereafter: Paper-II).

Compared with normal NLS1s, \rxj0134\ exhibits some highly unusual properties. Firstly, it exhibits dramatic X-ray variability, resembling an X-ray {\it complex} NLS1, but the spectrum becomes harder-when-brighter (\citealt{Grupe.2000}; \citealt{Komossa.2000}), which is not normally observed in AGN with high Eddington ratios (e.g. \citealt{Sobolewska.2009, Constantin.2009, Gu.2009, Emmanoulopoulos.2012, Soldi.2014}). Secondly, its radio-loudness was reported to be $R=71$ (\citealt{Grupe.2000}), consistent with a radio-loud (RL) classification for this NLS1, although we note that the radio detection in \cite{Grupe.2000} was made at $\sim 9$~GHz, rather than at $\sim 5$~GHz where the radio loudness index is defined. The combination of high radio-loudness and high Eddington ratio is also rare in the AGN population (e.g. \citealt{Yuan.2008, Yang.2020}). However, the upper limit to the {\it Fermi} $\gamma$-ray flux of \rxj0134\ is more than one order of magnitude fainter in X-rays than those seen in {\it Fermi} detected RL NLS1s (\citealt{Foschini.2015}). Also, our new monitoring campaign reveals the presence of a variable radio source with flux densities typical of Radio-Quiet (RQ) AGN. Hence it is not likely that the X-ray emission of \rxj0134\ is dominated by the presence of a jet. Its radio properties together with the long-term optical/UV/X-ray variability will be presented in a companion paper (Panessa et al. in preparation, hereafter: Paper-III).

A physical scenario proposed to explain the X-ray spectral-timing properties of \rxj0134\ involves obscuration, in which the 
observed X-ray variability results either from the variation of ionization state, or changes in gas column density along the line of sight (\citealt{Komossa.2000}; \citealt{Grupe.2000}). However, the short duration of the previous observations does not allow a more detailed analysis of the X-ray properties.

\subsection{The New Multi-wavelength Monitoring Campaign}
The unusual properties of \rxj0134\ motivated us to conduct a new multi-wavelength campaign on this exceptional NLS1 (principal investigator: C. Jin). This program comprised joint observations in 2019-12-19 with \xmm\ (\citealt{Jansen.2001}), \nustar\ (\citealt{Harrison.2013}), \swift\ (\citealt{Gehrels.2004}) and the 2.3-m optical telescope in the Siding Spring Observatory (SSO). These observations happened to coincide when \rxj0134\ was in its X-ray low-flux state exhibiting many interesting properties. Therefore, we triggered a follow-up multi-wavelength monitoring program on \rxj0134\ with target-of-opportunity observations using the Australia Telescope Compact Array (ATCA) and \swift\ satellite, in order to explore its long-term variability and possible state transition. The ATCA monitoring was conducted from 2020-01-31 to 2021-08-11 at 5.5, 9.0 and 18.0 GHz, with a cadence of 1 observation per 1-2 month. The \swift\ monitoring was conducted from 2020-01-31 to 2021-08-22\footnote{We have extended the \swift\ monitoring program while writing this paper. The additional data will be reported in future papers.}, with a cadence of 1 observation per $\sim$ 15 days, and an average exposure time of $\sim$ 1.5 ks each. To assist in our study, we also made use of multi-wavelength archival data of \rxj0134. All the datasets used in our work are listed in Table~\ref{tab-obs}.

\subsection{Scope of This Paper}
We will publish the complete results of this new multi-wavelength campaign in a series of papers. In this first paper, we report the results from our detailed X-ray spectral-timing analyses based on the simultaneous \xmm, \nustar\ and \swift\ observations in 2019-12-19. Archival X-ray observations of \rxj0134\ are also analyzed for comparison.

This paper is organized as follows. Firstly, we describe the X-ray datasets of \rxj0134\ and the data reduction procedures. Then we present the long-term drastic X-ray spectral variability in Section 3 . The detailed short-term X-ray spectral-timing properties are presented in Section 4, including light curves and power spectra in different energy bins, as well as frequency-differentiated rms, covariance and lag spectra. A full spectral-timing modelling is described in Section 5. We discuss the peculiar X-ray variability and its implications in Section 6, as well as the short-term ultraviolet (UV) variability revealed by \xmm, simultaneously. The final section summarizes the main results of this paper. Throughout the paper, we adopt a flat universe model with the Hubble constant H$_{0} = 72$ km s$^{-1}$ Mpc$^{-1}$, $\Omega_{\Lambda} = 0.73$ and $\Omega_{\rm M} = 0.27$.

\begin{table}
\centering
   \caption{The new multi-wavelength campaign on \rxj0134\ and the archival datasets used in this series of papers. $T_{\rm obs}$ is the total observing time. For \nustar\ the Earth occultations and South Atlantic Anomaly (SAA) passages have been excluded from $T_{\rm obs}$.}
     \begin{tabular}{@{}lccr@{}}
     \hline
    Instrument & Obs-Date & $T_{\rm obs}$ & Waveband \\
     & & (ks) &\\
    \hline
    \multicolumn{4}{c}{New Multi-wavelength Campaign} \\
    {\it NuSTAR}& 2019-12-19 & 98.3 & Hard X-ray \\
    \xmm\ & 2019-12-19 & 134.3 & X-ray/UVW1 \\
    {\it SSO Telescope} & 2019-12-19 & 1.8 & Optical \\
    {\it Swift} & from 2019-12-19 & & X-ray/UV/Optical \\
    & to 2021-08-22 & \multicolumn{2}{l}{51 obs of 0.2-2.9 ks each} \\
    {\it ATCA} & from 2020-01-31 &  & Radio (5.5, 9.0, 18.0 GHz) \\
    & to 2021-08-11 & \multicolumn{2}{l}{16 obs} \\
    \hline
     \multicolumn{4}{c}{Archival Observations} \\
    \xmm\ & 2008-12-11 & 32.1 & X-ray/UV/Optical \\
    \asca\ & 1997-12-10 & 41.4 & X-ray \\
    \rosat\ & 1996-06-06 & 7.2 & Soft X-ray \\
    {\it HST} (FOS) & 1996-09-21 & 1.7 & UV (G130H) \\
    {\it HST} (FOS) & 1996-09-21 & 2.1 & UV (G130H) \\
    {\it HST} (FOS) & 1996-09-21 & 0.2 & UV (G160L) \\
    {\it HST} (FOS) & 1996-09-21 & 1.5 & UV (G190H) \\
    {\it HST} (FOS) & 1996-09-21 & 1.2 & UV (G270H) \\
    {\it HST} (FOS) & 1996-09-21 & 1.0 & Optical (G400H) \\
    {\it HST} (FOS) & 1996-09-21 & 0.6 & Optical (G570H) \\
    {\it WISE} & 2010-06-20 & -- & Infrared (Band 1-4) \\
    {\it 2MASS} & 1999-08-27 & -- & Infrared (J, H, K)\\
    \hline
     \end{tabular}
\label{tab-obs}
\end{table}

\section{Observations and Data Reduction}
\label{sec-obs}
This new campaign of \rxj0134\ comprises deep simultaneous observations by \xmm\ and \nustar, whose results are reported in this work. We first describe the standard data reduction procedures below.

\xmm\ has observed \rxj0134\ twice. These two observations are separated by 11 years. We refer to the first \xmm\ observation in 2008 as XMM-1, and our new observation in 2019 as XMM-2. During both observations, the three European Photon Imaging Cameras (EPIC) were all set in the full-frame mode with thin filter. The data were downloaded from the \xmm\ Science Archive (XSA), and were reduced with the \xmm\ Science Analysis System (SAS v18.0.0) and the latest calibration files. The raw data from the European Photon Imaging Camera (EPIC) were reprocessed with the {\tt epproc} and {\tt empproc} tasks to produce event files. 

A circular aperture with a radius of 35 arcsec was used to extract source events, while background events were extracted from a nearby source-free region of a larger size. The effect of photon pileup was checked with the {\tt epatplot} task and was found to be negligible. For EPIC-pn, we also checked that both the source and background regions were not affected by the copper line contamination (\citealt{Katayama.2004}). We used the {\tt evselect} task to extract source and background light curves and spectra. The {\tt rmfgen} and {\tt arfgen} tasks were used to produce the response and auxiliary files for the spectra, and the {\tt epiclccorr} task was used to perform background subtraction and various corrections for the light curves. During XMM-2, the optical monitor (OM) was set in the {\sc Imaging+Fast} mode with the UVW1 filter. However, we find that the {\sc Fast} mode data was spoiled by an issue of satellite pointing accuracy{\footnote{Private communications with the \xmm\ helpdesk.}}, although the {\sc Imaging} mode photometry was not affected. During XMM-1, the OM was set in the {\sc Imaging} mode with five different filters, i.e. UVW2, UVM2, UVW1, U, B. We used the {\tt omichain} to reprocess the data and derive photometric fluxes.

This new campaign also includes the first \nustar\ observation of \rxj0134, which has a total exposure time of 98.3 ks. We used the tasks included in the HEASoft package (v6.27.2, \citealt{Blackburn.1995}) to process the data. The {\tt nupipline} task was used to reprocess the data with the latest calibration files. The data during the passages of the South Atlantic Anomaly (SAA) region were excluded by setting the parameter `saamode' to `optimized'. The {\tt nuscreen} task was used to create event files for the `grades' of 0-4. The {\tt nuproducts} task was used to produce the source and background spectra, response and auxiliary files. The source extraction region was chosen to be a circular region with 1 arcmin radius centered at the source coordinates, while background was extracted from a nearby source-free region with the same size.

\section{Time-average X-ray Spectra and Long-term Variability}
\label{sec-xray-spec}
\begin{figure}
\centering
\includegraphics[trim=0.1in 0.4in 0.0in 0.1in, clip=1, scale=0.51]{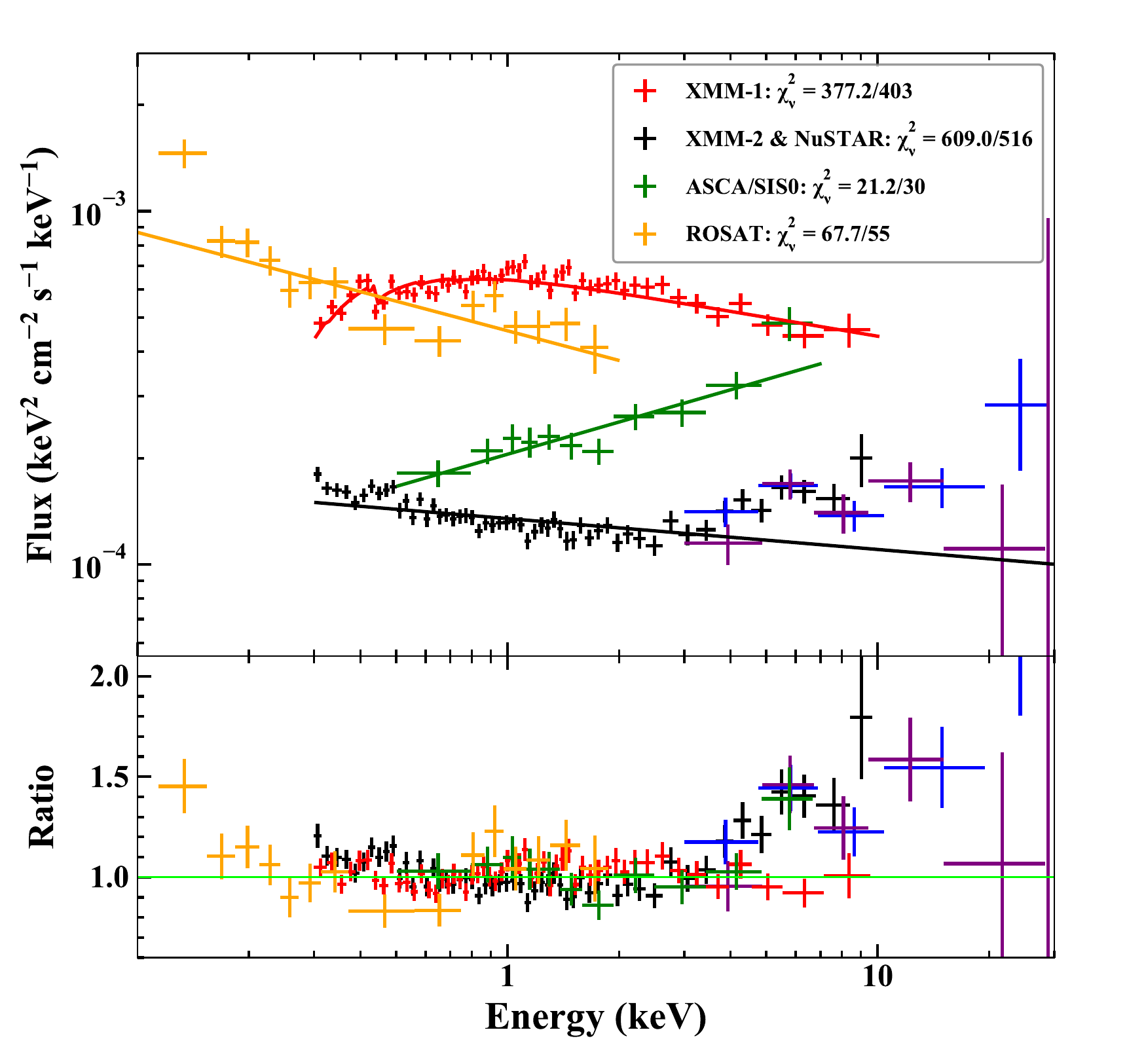} 
\caption{The X-ray spectra of \rxj0134\ observed by \xmm\ in XMM-1 (red) and XMM-2 (black). XMM-2 also has a simultaneous \nustar\ observation (FPMA: blue; FPMB: purple). The {\rosat} (orange) and {\asca} (green) spectra are also plotted for comparison. The solid lines indicate the best-fit absorbed power law models. All the spectra have been corrected for the Galactic absorption of $1.77\times10^{20}$ cm$^{-2}$. The soft X-ray excess is very weak or even undetected, and the spectral variability is remarkable. Significant excess flux is also observed above 4 keV in XMM-2 and the simultaneous \nustar\ observation, but not in XMM-1.}
\label{fig-xrayspec}
\end{figure}

First we checked the long-term X-ray variability of \rxj0134\ by comparing the latest observations with archival observations.
Figure~\ref{fig-xrayspec} shows the X-ray time-average spectra from our new XMM-2 and \nustar\ dataset (black/blue/purple), compared to the X-ray spectra from \asca\ (green), XMM-1 (red) and \rosat\ (yellow), which were observed 10-20 years ago (see Table~\ref{tab-obs}). These spectra have all been corrected for the Galactic column density of $1.77\times10^{20}$~cm$^{-2}$ (\citealt{Willingale.2013}), using the {\tt tbabs} model in {\sc xspec} (v12.11.0k, \citealt{Arnaud.1996}). We adopt the cross-sections taken from \citet{Verner.1996} and abundances taken from \citet{Wilms.2000}. There are drastic changes in both the X-ray flux and spectral shape among these observations.

We then fit all the spectra with a single power law, allowing for the possibility of intrinsic absorption in the host galaxy using {\tt ztbabs}. The archival \rosat\ 
spectra are quite well fit with a steep power law, with the
best-fit photon index being $\Gamma~=~2.28\pm0.07$ and no 
intrinsic absorption ($\chi^2_\nu=67.5/55$: orange in Figure~\ref{fig-xrayspec}), though the spectrum can be 
better fit including an ultra-soft low temperature component below 
0.2~keV (\citealt{Komossa.2000}). This is in sharp 
contrast to the much harder \asca\ spectrum, which has $\Gamma~=~1.70$ $\pm$ $0.08$, but again with no intrinsic absorption (green in Figure~\ref{fig-xrayspec}). A quite steep spectrum is seen again in XMM-1, with $\Gamma~=~2.18$ $\pm$ $0.03$, but this time with significant intrinsic absorption, $N_{\rm H}={(4.40\pm0.91)}\times10^{20}$ cm$^{-2}$ ($\chi^2_\nu=377.2/483$: red in Figure~\ref{fig-xrayspec}). The new spectrum in XMM-2 (black) is a factor of $\sim$ 5 fainter than in XMM-1 below 4~keV. Then it shows a pronounced rise to higher energies, which is also confirmed by the \nustar\ spectra. The power law fit gives
 $\Gamma~=~2.09$ $\pm$ $0.02$, with no intrinsic absorption but 
it is clear that this still leaves excess flux below 0.5 keV and above 4 keV.

\begin{figure*}
\centering
\begin{tabular}{cc}
\includegraphics[trim=0.0in 0.0in 0.0in 0.1in, clip=1, scale=0.51]{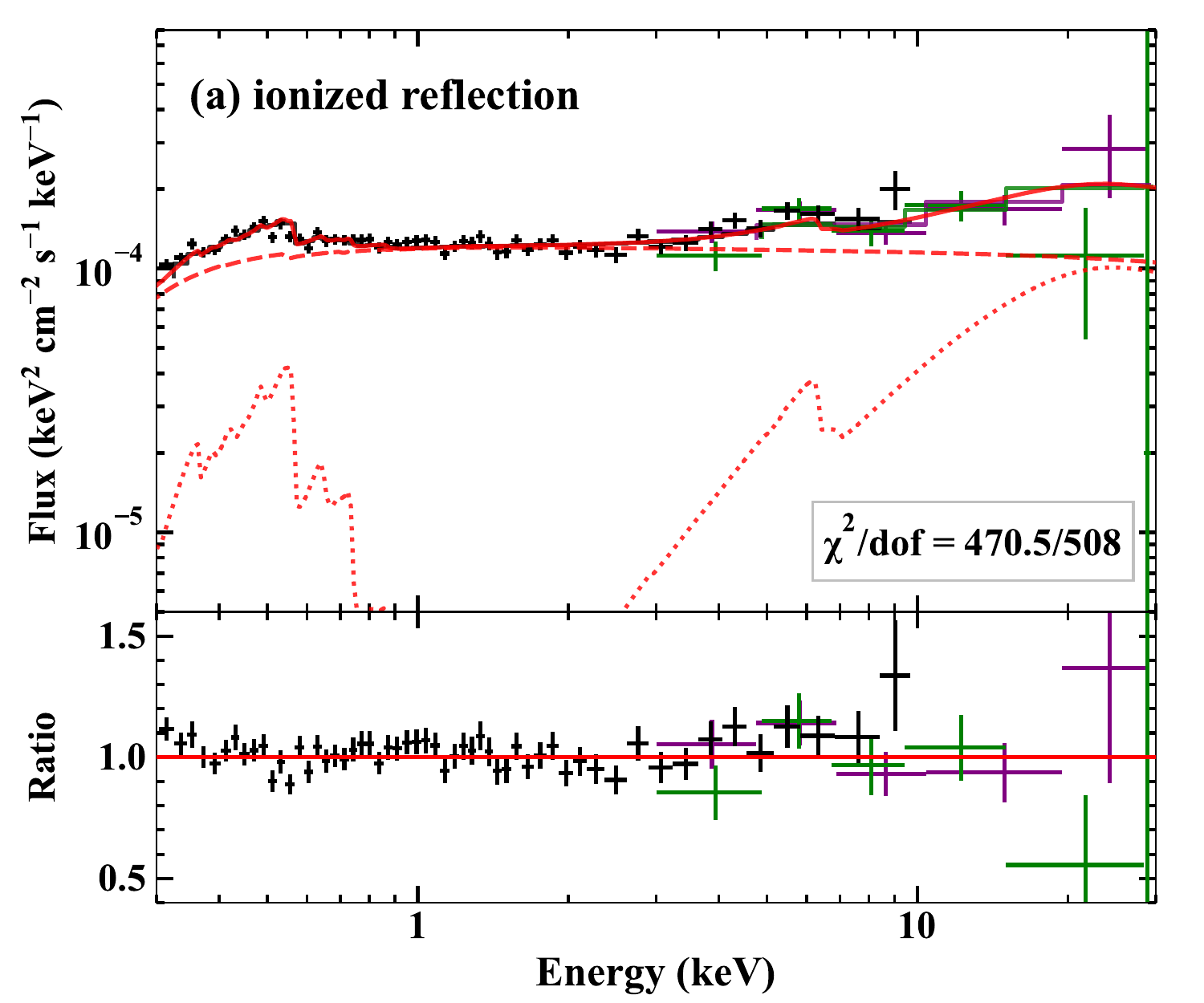} &
\includegraphics[trim=0.0in 0.0in 0.0in 0.1in, clip=1, scale=0.51]{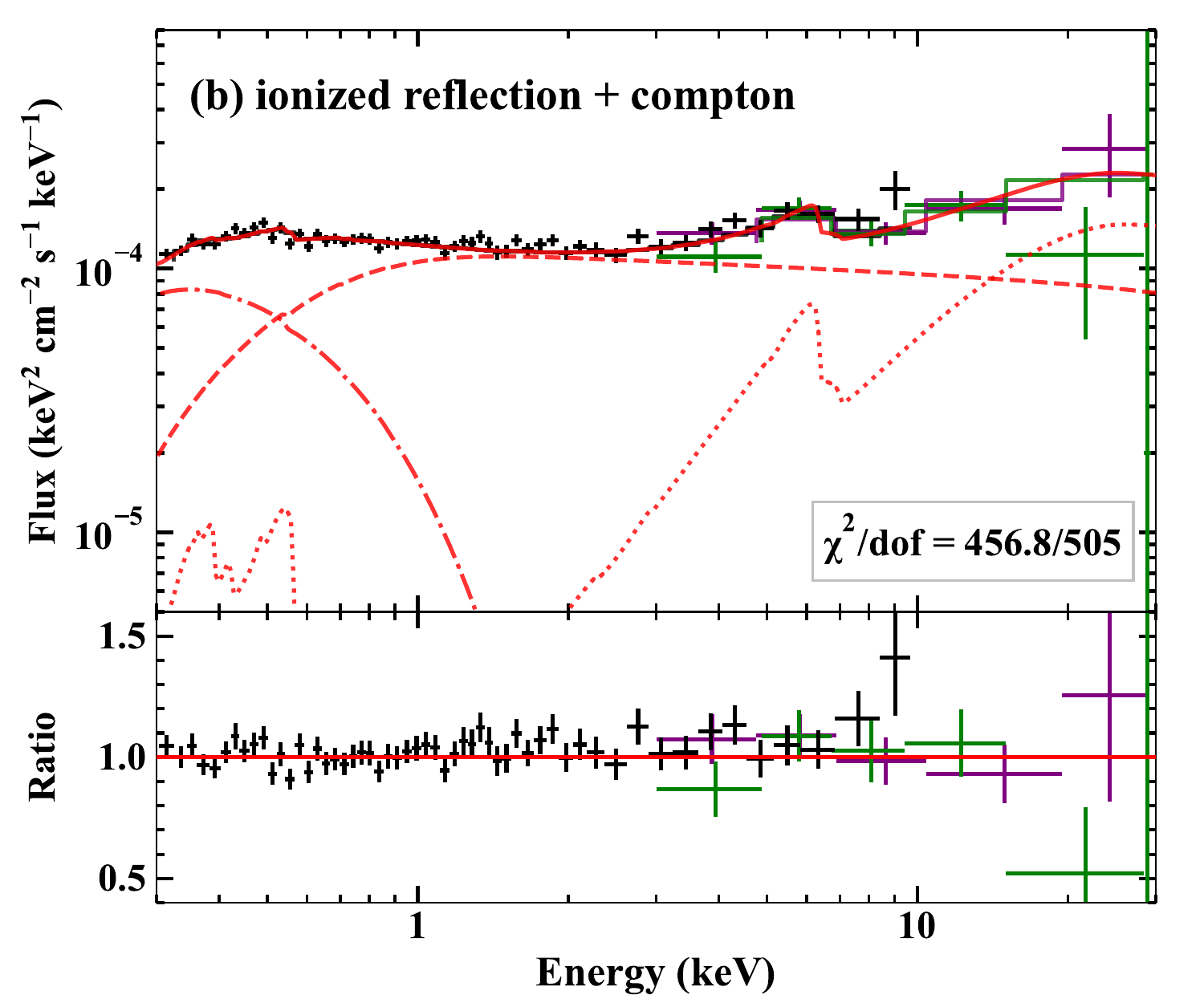} \\
\includegraphics[trim=0.0in 0.0in 0.0in 0.1in, clip=1, scale=0.51]{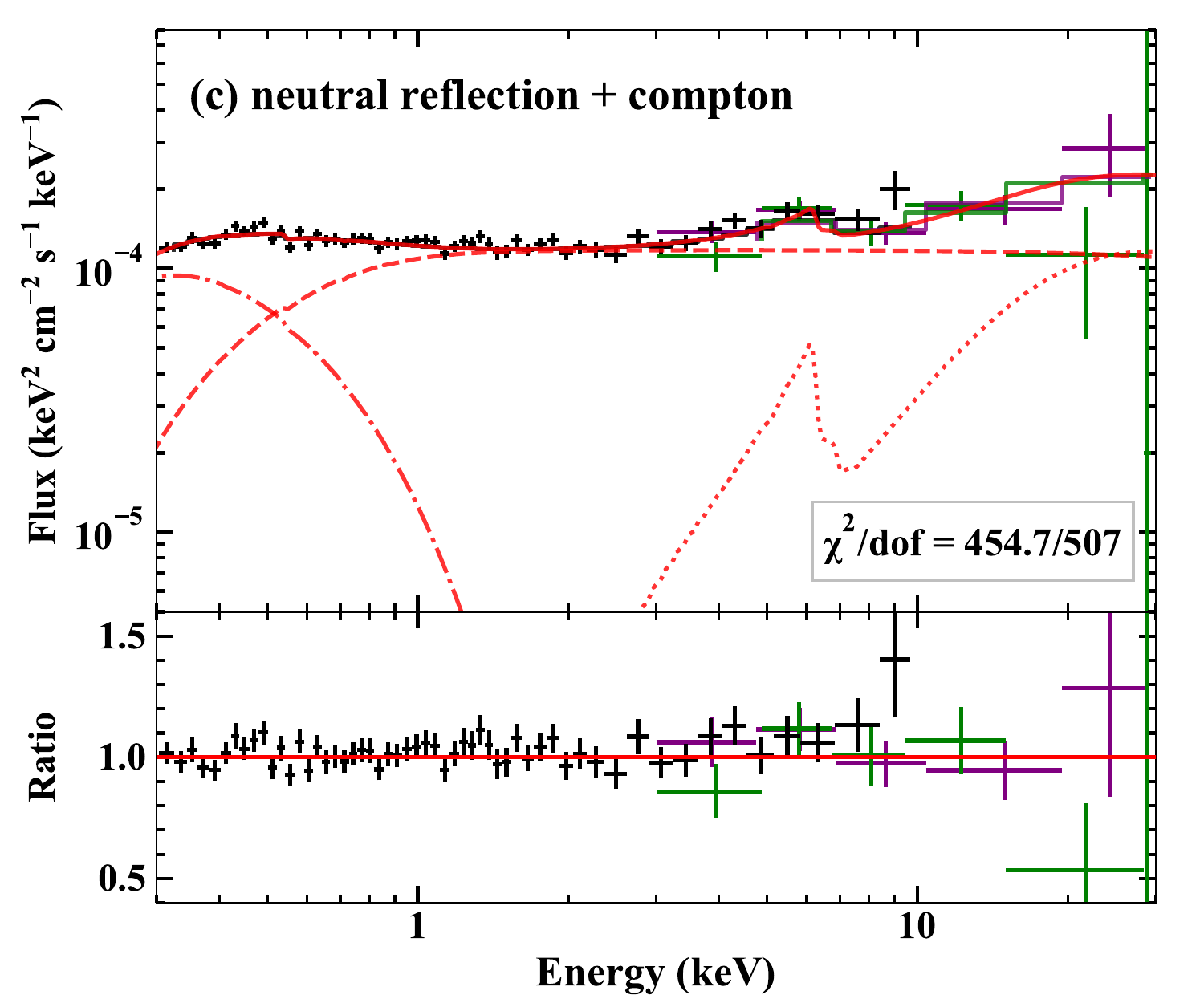} &
\includegraphics[trim=0.0in 0.0in 0.0in 0.1in, clip=1, scale=0.51]{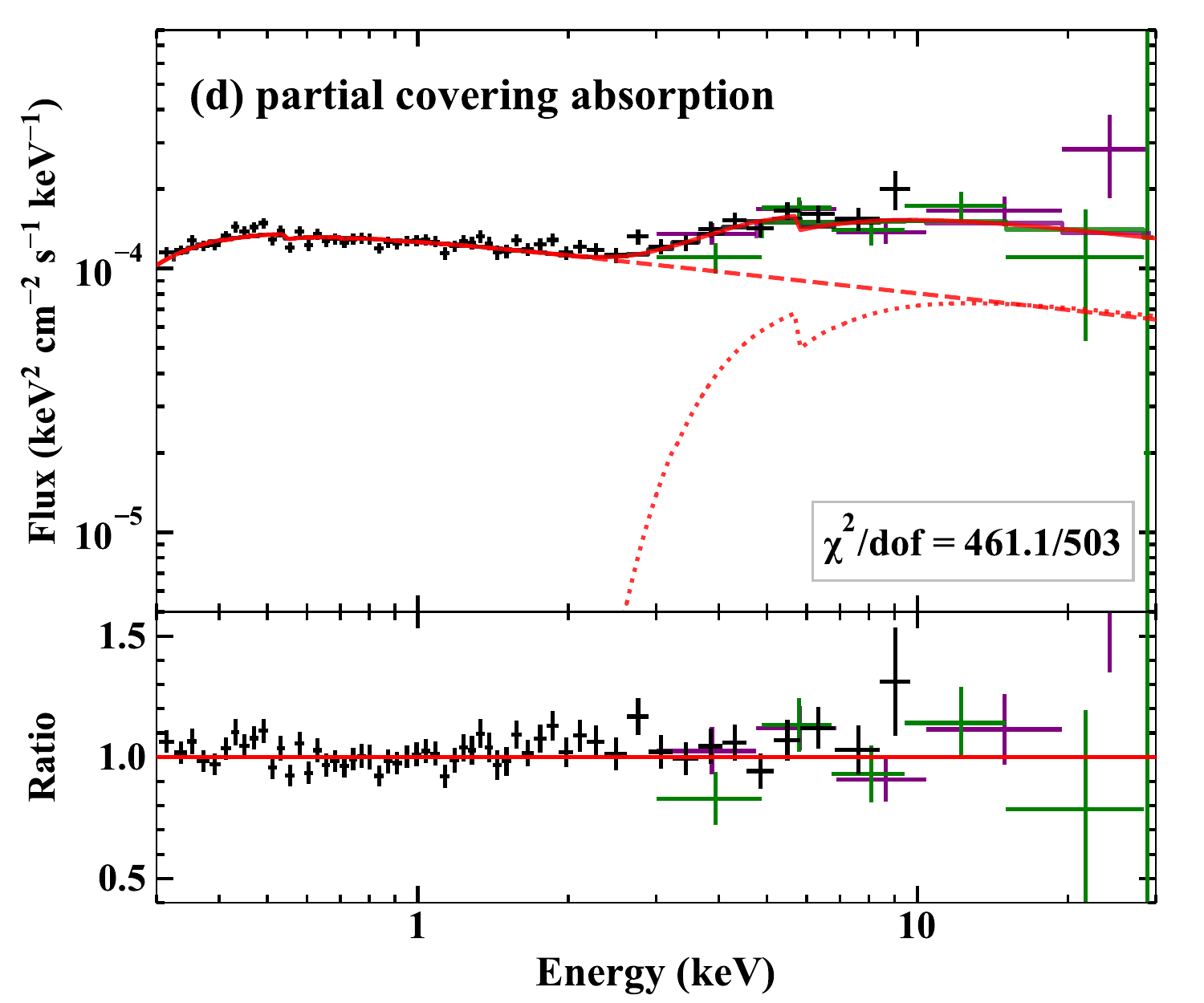} \\
\end{tabular}
\caption{Fitting the \xmm\ and \nustar\ spectra of \rxj0134\ observed in XMM-2 with reflection and absorption models. Panel-a: unfolded spectra based on the best-fit ionized reflection model {\tt relxill} (red dotted line). The incident continuum is a power law (red dash line). Panel-b: adding a soft X-ray Comptonisation (red dash-dot line) to {\tt relxill}. The incident continuum is a hard Comptonisation {\tt nthcomp}. Panel-c: replacing {\tt relxill} with the neutral reflection model {\tt pexmon}. Panel-d: the partial covering absorption model. All the above models can reproduce the hard X-ray excess above 4 keV.}
\label{fig-specfit1}
\end{figure*}

\begin{figure*}
\centering
\begin{tabular}{cc}
\includegraphics[trim=0.2in 0.2in 0.0in 0.1in, clip=1, scale=0.5]{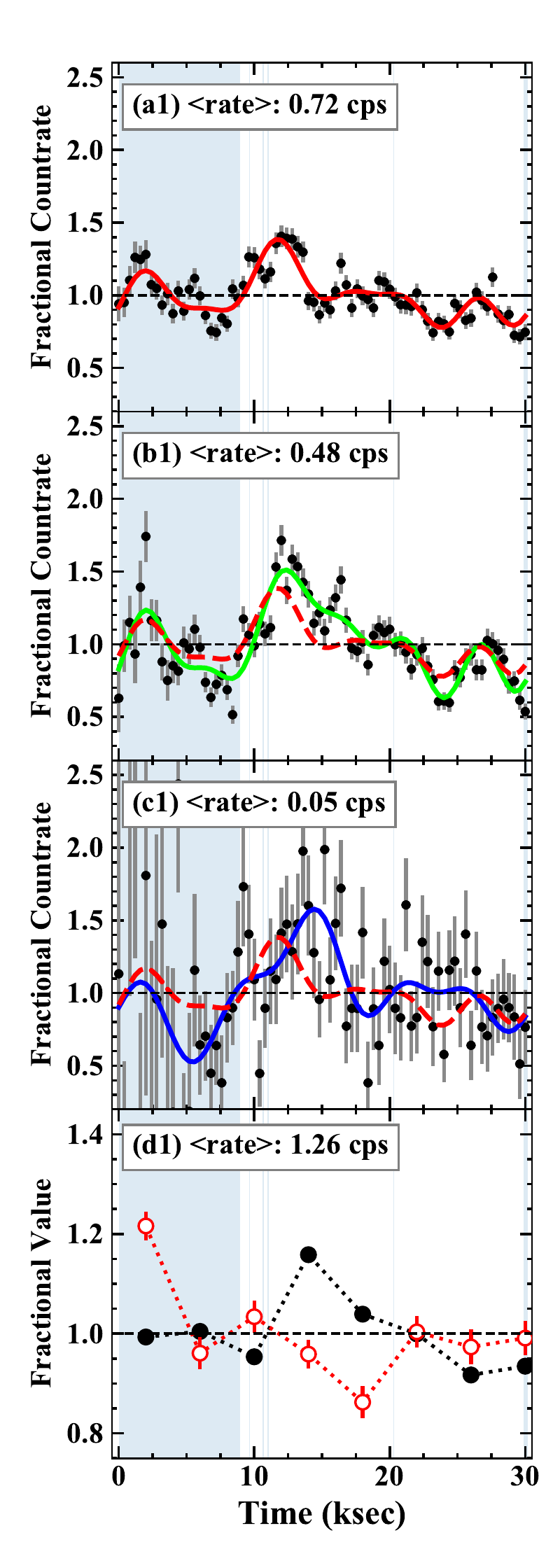} &
\includegraphics[trim=0.3in 0.2in 0.0in 0.1in, clip=1, scale=0.5]{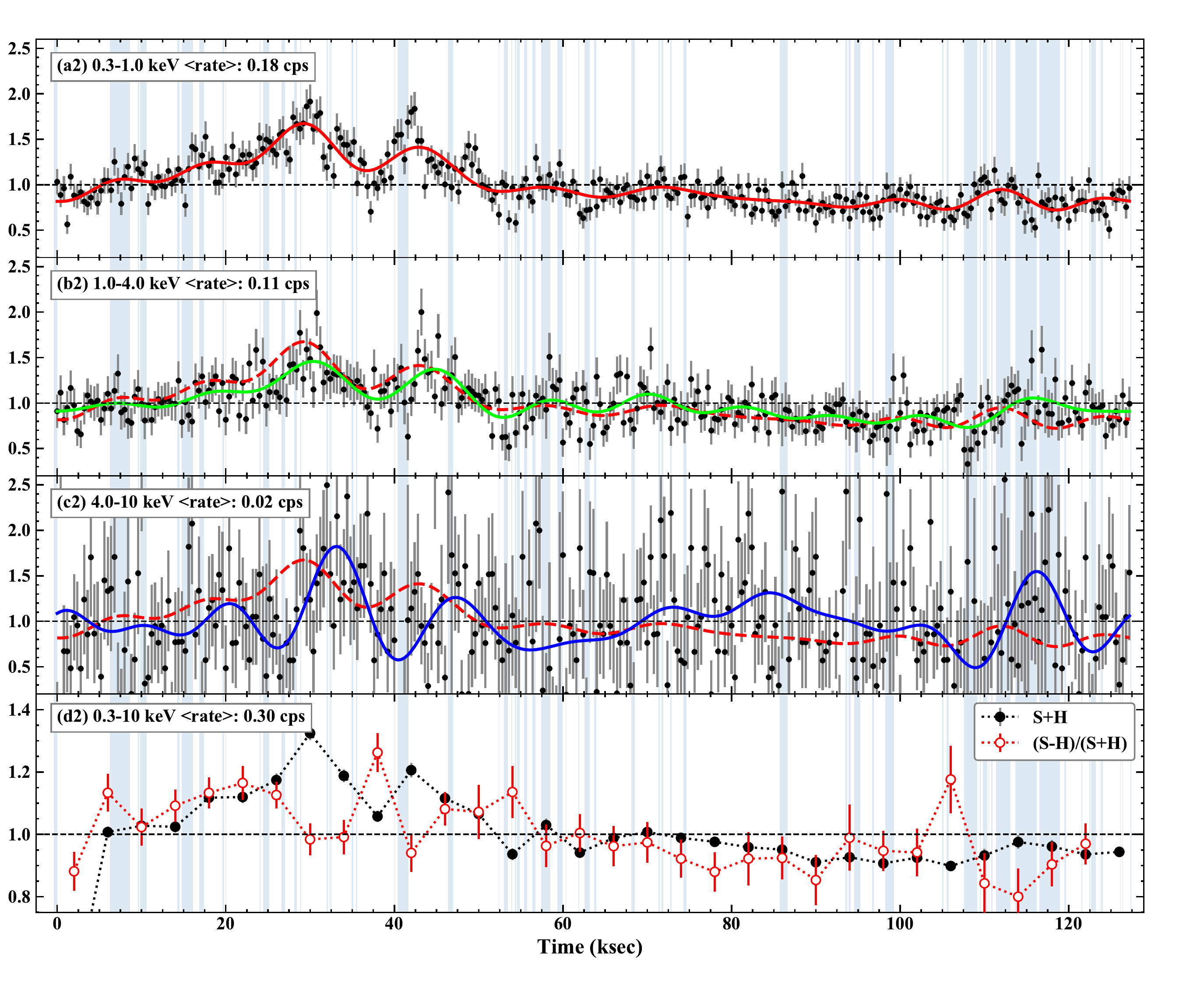} \\
\end{tabular}
\caption{Short-term X-ray variability of \rxj0134\ in \xmm\ XMM-1 (Panels a1-d1) and XMM-2 (Panels a2-d2). The top three rows show the 400 s binned light curves in 0.3-1, 1-4 and 4-10 keV bands in the fractional unit. In each of these panels, the solid line shows the source variability after applying a low-pass filter to the light curve. The filter is $\le 2\times10^{-4}$ Hz for XMM-1 and $\le 10^{-4}$ Hz for XMM-2. In the second and third rows, the red dash lines also show the low-pass filtered light curve in 0.3-10 keV for comparison. In the fourth row, the 0.3-10 keV light curve is plotted in black, which is binned with 4 ks and its amplitude is scaled down by 0.5 for clarity. The variation of spectral shape is plotted in red, which is defined as $(S-H)/(S+H)$, where $S$ and $H$ stand for 0.3-1.5 keV and 1.5-10 keV. The vertical light-blue regions indicate the time intervals where background flares are detected.}
\label{fig-lc}
\end{figure*}

\begin{figure*}
\centering
\includegraphics[trim=0.0in 0.0in 0.0in 0.0in, clip=1, scale=0.46]{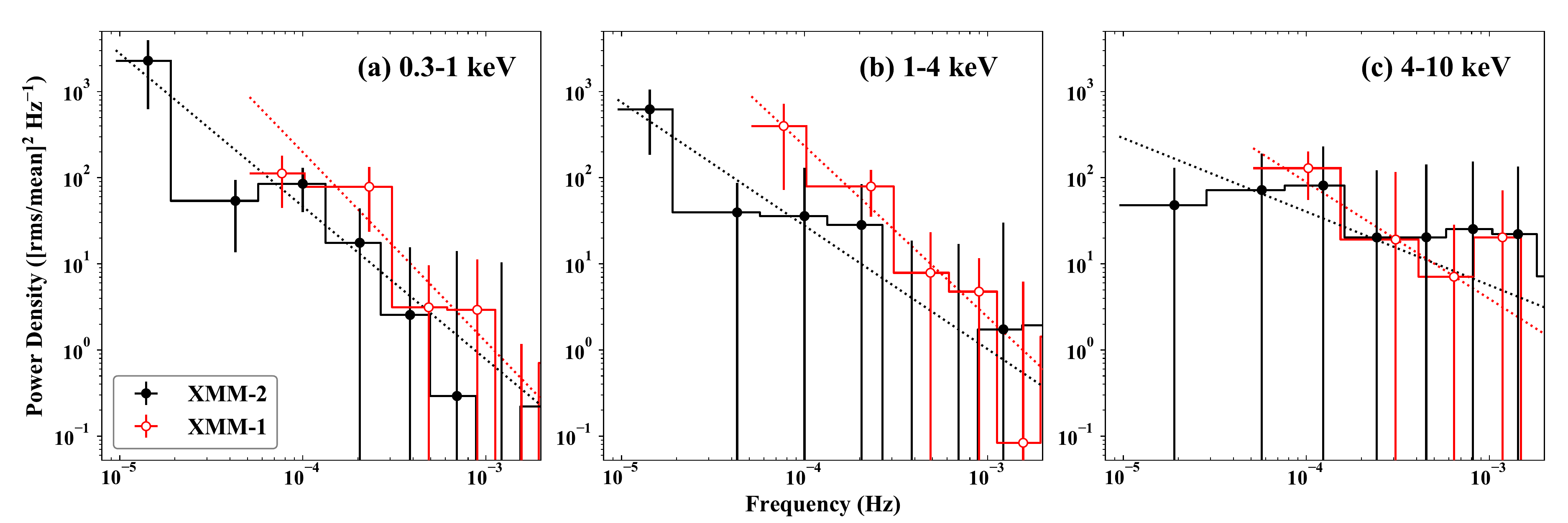} \\
\caption{The X-ray PSDs of \rxj0134\ in 0.3-1, 1-4 and 4-10 keV as observed by \xmm\ EPIC-pn in XMM-1 (red) and XMM-2 (black). Poisson noise power has been subtracted. The dotted lines indicate the best-fit power law models for the continuum noise.}
\label{fig-psd}
\end{figure*}

The higher energy excess observed in XMM-2 looks like either reflection or a heavily absorbed component. We first assume that this is due to reflection, and fit it with the {\tt relxill} model (\citealt{Garcia.2013}). The incident continuum is assumed to be a single power law. Since the spectra are of limited quality, we fix the relativistic smearing parameters
to illumination of $3$, spin 0, $R_{\rm in}~=~6~R_{\rm g}$, $R_{\rm out}~=~400~R_{\rm g}$ and inclination of 60$^\circ$. The model gives $\chi^2_\nu~=~470.5/508$ with an
ionization parameter of $\log\xi~=~0.84$. This low ionization allows {\tt relxill} to contribute a small amount of flux below 0.8 keV (see Figure~\ref{fig-specfit1}a). However, if we add a Comptonisation component ({\tt compTT}: \citealt{Titarchuk.1994}) to the soft X-ray band, and replace the power law with a hard X-ray Comptonisation ({\tt nthcomp}: \citealt{Zdziarski.1996, Zycki.1999}), the soft X-rays below 0.5 keV are then dominated by {\tt compTT} (see Figure~\ref{fig-specfit1}b). The ionization parameter of {\tt relxill} pegs at its lower limit of $\log\xi~=~0$, and the $\chi^2$ decreases to 456.8 for 505 degrees-of-freedom (dof), both indicating that the soft X-ray spectrum favors the smoother component shape of {\tt compTT}. We also see that even at the lowest possible ionization of the {\tt relxill} model, there is still a small reflected contribution at low energies.

Then we replace {\tt relxill} with {\tt pexmon} (\citealt{Nandra.2007}), which is a neutral reflection model and produces various line features. Since the reflection may happen at tens to hundreds of $R_{\rm g}$ (see Section~\ref{sec-discussion1}), it is also necessary to consider the relativistic effect. Besides, despite the low signal-to-noise above 4 keV, the hard X-ray excess appears rather smooth, which also suggests the necessity of smearing the line features produced by {\tt pexmon}. Therefore, we include the relativistic effect with the {\tt kdblur} model (\citealt{Laor.1991}). Firstly, the parameters of {\tt kdblur} are kept the same as in {\tt relxill}. The best-fit result is shown in Figure~\ref{fig-specfit1}c. In this case the $\chi^2$ decreases to 454.7 for 507 dof, confirming that the spectra observed in XMM-2 and \nustar\ prefer neutral reflection rather than ionized reflection. Secondly, we allow the inner radius parameter $R_{\rm in}$ of {\tt kdblur} to be a free parameter, then the $\chi^2$ decreases slightly to 454.3, and the best-fit $R_{\rm in}$ is found to be $4.00^{+9.25}_{-2.76}~R_{\rm g}$. Thus it confirms that the spectra also prefer the neutral reflection component to be relativistically smeared.

However, we also find that the high energy excess above 4 keV is similarly well fitted
with an additional absorbed power law component, as shown in Figure~\ref{fig-specfit1}d, giving 
$\chi^2=461.1$ for 503 dof for a power law index tied to 
the power law which dominates at lower energies with $\Gamma~=~2.21\pm0.03$, absorbed by a column of $N_{\rm H}=3.69^{+1.40}_{-1.07}\times 10^{23}$~cm$^{-2}$. This model can be interpreted within the partial covering scenario. If we assume that there are no other types of obscurers, then the covering fraction, as determined from the ratio of normalization between the two power law components, can be found to be $51 \pm 5$ per cent. This means that 51 per cent the primary power law continuum is absorbed by this additional gas column, and the absorbed power law only contributes to the hard X-rays. We emphasize that both reflection and absorption models result in a similarly steep power law with $\Gamma~=~2.1-2.2$, with weak/no sign of intrinsic soft X-ray excess.

Our results confirm the long-term drastic variability of the X-ray flux and spectral shape of \rxj0134. There is no systematic harder-when-brighter behaviour for the long-term spectral variability. The harder-when-brighter behaviour was observed {\em within} the \rosat\ data, and is consistent with the steep power law varying on short timescales, while the strong ultra-soft component below 0.2~keV remained constant (\citealt{Grupe.2000}; \citealt{Komossa.2000}). The lack of a systematic pattern, and the clear evidence for variable absorption/reflection in the \xmm\ data make it probable that the long timescale variability is dominated by variable absorption along our line of sight. This explanation can also be extended to the very hard power law seen in the \asca\ data. Such a power law could be consistent with a jet, but the flux level is suspiciously in between the maximum and minimum seen from the intrinsically steep spectra seen in all other datasets, whereas it might be expected to be brighter if the jet adds to (and dominates) the accretion flow emission. Instead, this spectrum could be the result of absorption and/or reflection in a more complex configuration than that seen in XMM-2.

None of the spectra of \rxj0134\ display a classic soft X-ray excess as seen in other 
super-Eddington NLS1s. The spectra of these typical NLS1s generally start to deviate from the hard X-ray power law below 2 keV, and then their data-to-model ratios reaches at least a factor of few at 0.3 keV to form typical soft excesses (e.g. \citealt{Jin.2009, Jin.2013, Jin.2016, Jin.2017a, Kara.2017, Parker.2018, Petrucci.2018}). The soft excess is also seen in RL NLS1 where the hard X-rays are dominated by the jet (e.g. 1H 0323+342, \citealt{Kynoch.2018}). In comparison, the data-to-model ratio of \rxj0134, as shown in Figure~\ref{fig-xrayspec}, is only $\lesssim$ 1.2 at 0.3 keV. Hence the weak soft excess makes \rxj0134\ a rather special NLS1.

Moreover, the hard X-ray excess can be fit equally well using reflection and absorption models. Hence 
it is necessary to examine the short-term X-ray variability in order to break the model degeneracy (e.g. \citealt{Jin.2021}), which is described in the next section.

\begin{figure}
\centering
\includegraphics[trim=0.0in 0.2in 0.0in 0.0in, clip=1, scale=0.56]{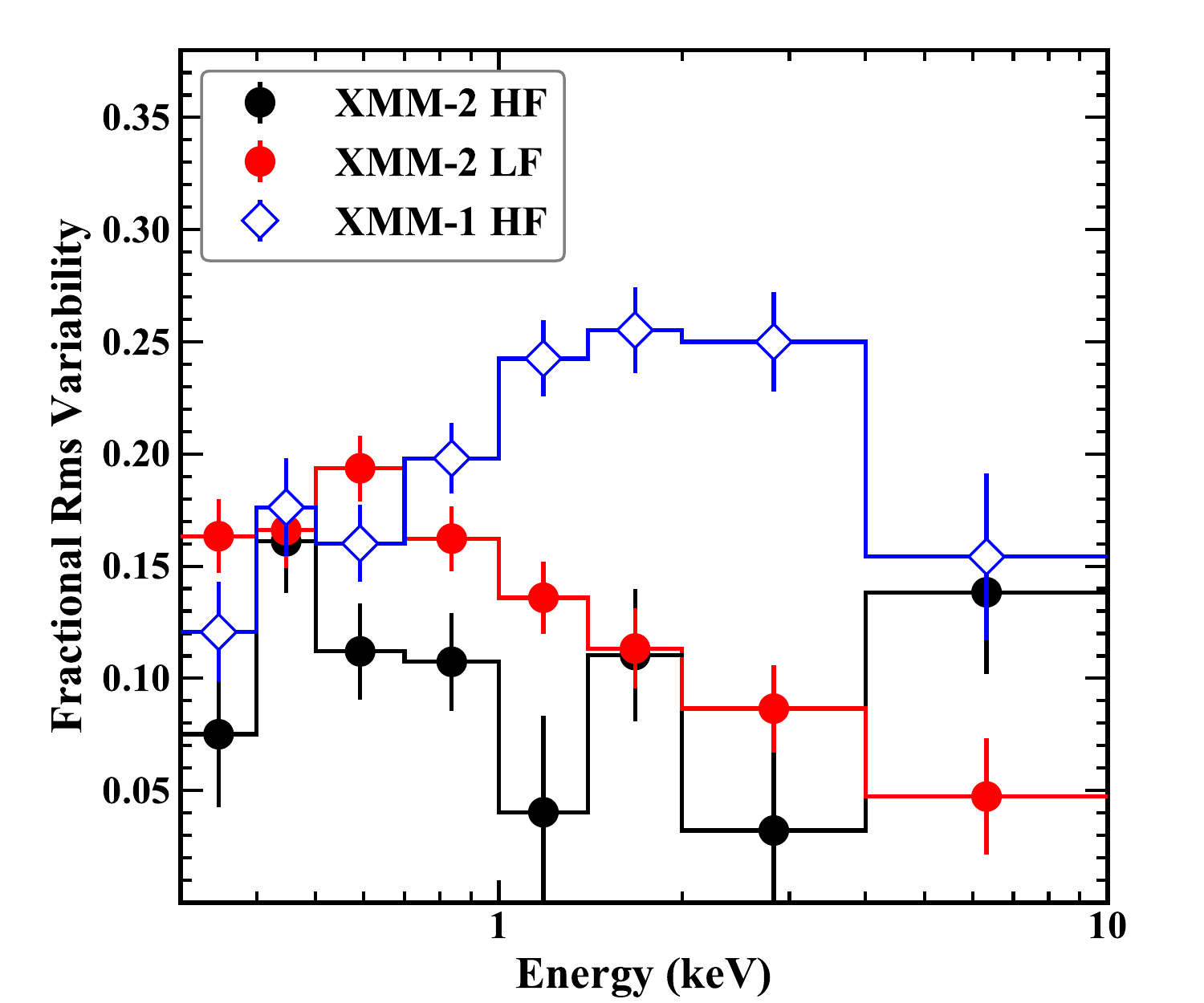} 
\caption{Fractional rms spectra of \rxj0134\ in different frequency bands. The black and red data points indicate the high-frequency (HF: $[0.05,1]\times10^{-3}$ Hz) and low-frequency (LF: $[0.95,5]\times10^{-5}$ Hz) rms spectra in XMM-2. The blue data indicate the HF rms spectrum in XMM-1.}
\label{fig-rmsspec}
\end{figure}

\begin{figure*}
\centering
\includegraphics[trim=0.0in 0.1in 0.0in 0.0in, clip=1, scale=0.7]{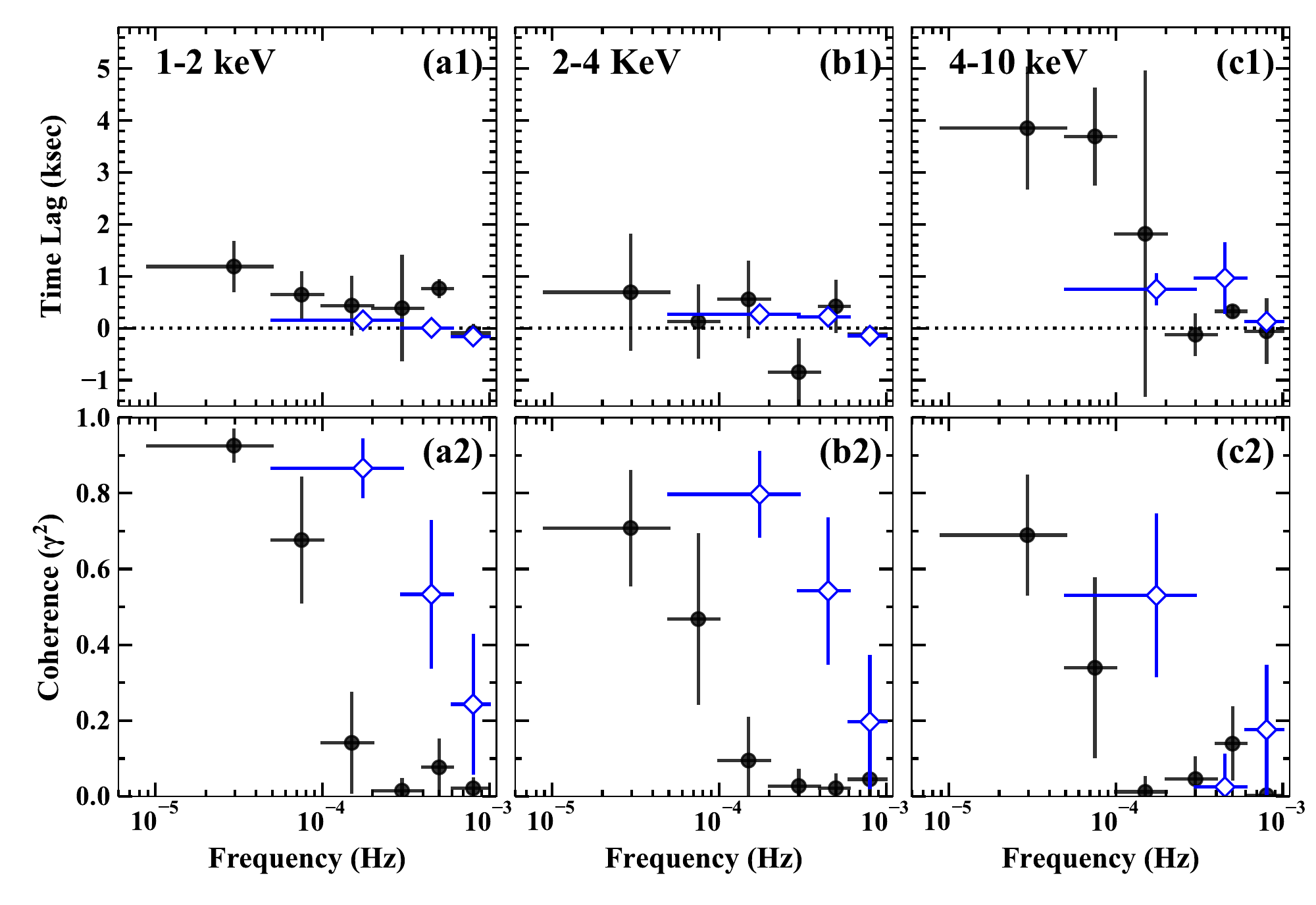} 
\caption{The frequency-differentiated time lag and coherence in three energy bands. The reference band is 0.3-1 keV. The blue and black data points are for XMM-1 and XMM-2 separately.}
\label{fig-cohlag}
\end{figure*}

\begin{figure}
\centering
\includegraphics[trim=0.0in 0.1in 0.0in 0.1in, clip=1, scale=0.55]{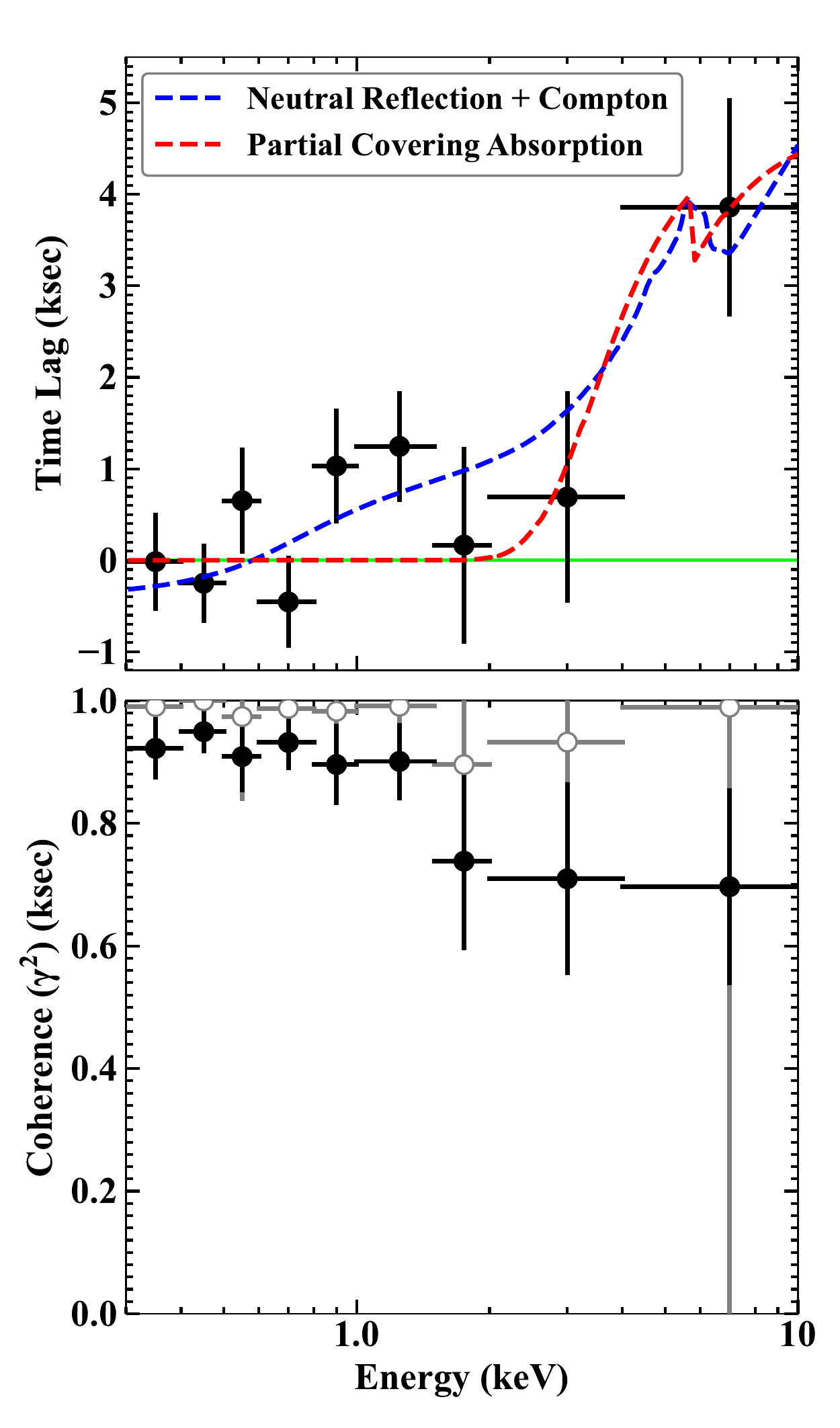} \\
\caption{Upper panel: the lag spectrum relative to 0.3-1 keV in $[0.95,5]\times10^{-5}$ Hz. The blue dash line indicates the best-fit lag model based on the best-fit neutral reflection plus Comptonisation model as shown in Figure~\ref{fig-specfit4}, while the red dash line indicates the best-fit partial covering absorption model. Lower panel: the coherence spectrum relative to 0.3-1 keV, with the black and grey data points being the raw and Poisson-noise corrected coherences, respectively. All the coherences are all close to unity, so the time lags are statistically meaningful.}
\label{fig-lagspec2}
\end{figure}

\section{Short-term X-ray Variability}
\label{sec-xray-var}
\subsection{X-ray Light Curves}
We explore the nature of the \xmm\ spectra in more detail using the fast variability observed within each observation. The \nustar\ data are fragmented due to the significant orbital overheads, and so do not have good signal-to-noise due to the low count rates of \rxj0134. Thus we only use the \xmm\ data for variability study.

Figure~\ref{fig-lc} shows the background-subtracted light curves of \rxj0134, which are observed by \xmm\ EPIC-pn in XMM-1 and XMM-2. In XMM-1, the first 10 ks is contaminated by background flares, and only the later 20 ks is clean. In XMM-2, short-period background flares spread in the entire observing window, as shown by the shaded regions in Figure~\ref{fig-lc}. These flaring data gaps should be treated carefully during the variability study, which we describe in detail below. In order to compare variability between soft and hard X-ray bands, we also divide the energy band into 0.3-1~keV, 1-4 and 4-10 keV bands. We rebin the light curves with 400 s per bin to suppress the Poisson noise fluctuation. In XMM-1, the mean count rate in EPIC-pn is 0.72, 0.48 and 0.05 counts per second (cps) in the three energy bands. In XMM-2, the mean count rate decreases to 0.18, 0.11 and 0.02 cps, respectively.

To visualize the intrinsic variability and the inter-band correlation, we apply a low-pass filter to the light curves. The filter is $\le 2\times10^{-4}$ Hz for XMM-1 and $\le 10^{-4}$ Hz for XMM-2, and the results are shown as solid lines in Figure~\ref{fig-lc}. The red dash lines in the second and third rows both show the low-pass filtered light curve in 0.3-1 keV (i.e. identical to the red solid line in the first row), which are used for comparison with the variability in 1-4 and 4-10 keV.
Significant variability is detected across the entire 0.3-10 keV band, which is also accompanied by significant and complex inter-band correlations. In addition, in order check the short-term variability of the spectral shape, we produce the time series of the hardness ratio, which is shown in Figure~\ref{fig-lc}d. The soft band is defined as 0.3-1.5 keV, and the hard band is defined as 1.5-10 keV. It can be seen that there is a tentative softer-when-brighter behavior on these short timescales in XMM-2, although the Pearson's correlation coefficient is only 0.29, and no such behavior is seen in XMM-1. The variability shown in these light curves allows us to conduct more detailed spectral-timing analysis in the frequency domain.

\subsection{X-ray Power Spectral Density}
\label{sec-psd}
As a further comparison of variability between different energy bands, we calculate the power spectral density (PSD)\footnote{Strictly speaking, this should be called {\it periodogram}, but in this paper we will call it PSD for simplicity.} of the light curves 0.3-1, 1-4 and 4-10 keV bands. To reduce the influence of data segments contaminated by background flares, we exclude all the data after 105 ks in XMM-2. This allows us to explore the PSD down to $\sim10^{-5}$ Hz. We also exclude all the background contaminated short segments, and then fill these gaps with the mean count rate randomized by the Poisson noise (\citealt{Gonzalez.2012, Alston.2015, Ashton.2021}). As shown in Figure~\ref{fig-lc}, the length of these data gaps spans from 50 s to 2.4 ks, and their average length is only 427 s, thus they only affect the variability in the frequency band $\gtrsim10^{-3}$ Hz, which is dominated by the Poisson noise.

Then we use a single power law plus a free constant to model the red noise and Poisson noise, and derive the best-fit model with the maximum likelihood estimate method (\citealt{Vaughan.2010}). Figure~\ref{fig-psd} shows the Poisson noise subtracted PSDs in different energy bands, along with the best-fit power law model as indicated by the dotted lines. The variability of \rxj0134\ is significantly detected at $\lesssim 2\times10^{-4}$ Hz, which is much lower than the frequency band that might be affected by the subtraction of background flares. This figure also shows that the soft X-rays exhibit stronger low-frequency variability than the hard X-rays, which is a common phenomenon observed in other super-Eddington NLS1s (e.g. \citealt{Jin.2013, Jin.2016, Jin.2017a}). In XMM-2, the high-frequency band above $\gtrsim10^{-4}$ Hz is mostly dominated by the Poison noise power.

XMM-1 has only one clean data segment of 20 ks long, and so the PSD only extends down to $5\times10^{-4}$ Hz. We plot these PSDs in Figure~\ref{fig-psd} in red. Since \rxj0134\ is brighter in XMM-1, the Poisson noise power is much lower at $\gtrsim 10^{-4}$ Hz. The low-frequency power in XMM-1 appears stronger than in XMM-2, suggesting that \rxj0134\ shows stronger variability in XMM-1. However, we note that the red noise leak might have a significant effect in XMM-1 due to the limited length of exposure time, so this should be explored by future observations with longer duration.

\subsection{X-ray Rms Spectra}
\label{sec-rms}
In order to quantify in more detail how the variability amplitude changes as a function of X-ray photon energy, we calculate the rms spectra in different frequency bands (\citealt{Uttley.2014}). To produce these spectra, we divide the 0.3-10 keV band into smaller energy bins, extract EPIC-pn light curves in these bins, exclude all the background contaminated intervals, and then fill them with the mean count rate randomized by the Poisson noise. We also rebin all the light curves with 500 s per bin to ensure that there is no zero-count time bin. Since the gap-filling method involves randomization, we generate 1000 light curves for subsequent analyses, in order to ensure that all the results are statistically stable and robust.

For each light curve, we follow the standard prescription in \citet{Vaughan.1997} and \citet{Arevalo.2008} to calculate the rms and errors. Then the mean rms is derived from all the light curves. As these light curves are binned with 500 s, the Nyquist frequency is $10^{-3}$ Hz. The lowest frequency is set by the length of each light curve, which is 20 ks for XMM-1 and 105 ks for XMM-2 after the exclusion of background flares. Hence, the lowest frequency is $5\times10^{-5}$ Hz for Obs-1 and $9.5\times10^{-6}$ Hz for Obs-2. We define $[0.05,3]\times10^{-3}$ Hz as the high-frequency (HF) band, and $\le 5\times10^{-5}$ as the low-frequency (LF) band. This allows us to compare the HF rms spectra between XMM-1 and XMM-2, and check the shape of the LF rms spectrum in XMM-2. 

Figure~\ref{fig-rmsspec} shows the rms spectra. 
XMM-1 shows the typical shape of HF rms spectra seen in normal X-ray {\it simple} NLS1s, where the rms of the fast variability is somewhat suppressed below 1~keV due to the low energy spectrum being dominated by a more slowly variable soft X-ray excess (\citealt{Jin.2009, Jin.2013, Jin.2017a, Jin.2021}).
This could indicate that while the time-average spectrum in XMM-1 does not show any clear evidence for a soft X-ray excess, there is actually some contribution from this component in the soft X-rays. Alternatively,
the suppression of variability occurs in the region where the spectrum starts to become absorbed, so it may instead indicate a contribution from a less variable, scattered/reflected power law component at low energies. In addition, the LF rms spectrum also shows a typical shape seen in normal X-ray {\it simple} NLS1s, where the variability increases towards soft X-rays (\citealt{Jin.2009, Jin.2013, Jin.2017a, Jin.2021}). Therefore, irrespective of which process occurs, it is clear that the time-average spectra in XMM-1 contains multiple components with different variability patterns, although its shape is statistically consistent with a single absorbed power law.

The HF rms in XMM-2 is significantly lower than in XMM-1 
at all energies, but especially between 1-4 keV where the power law dominates in all models. 
This is consistent with
a model where the 1-4 keV spectrum in XMM-2 is scattered rather than seen directly, so that the fastest variability is suppressed. The somewhat stronger variability below 1 keV and above 4~keV could be where more of the direct component is seen. But as we mentioned before, the HF rms in XMM-1 can be severely affected by the red noise leak. We can directly test these different models by examining the causal relationships in the X-ray variability across the energy band by constructing coherence and lag spectra.

\subsection{X-ray Coherence and Lag Spectra}
\label{sec-cohlag}
The significant X-ray variability of \rxj0134\ allows the frequency-resolved coherence and lag spectra to be produced (see \citealt{Uttley.2014} and references therein). These variability spectra are a powerful tool to constrain causal relationships between different spectral components (e.g. \citealt{Jin.2013}; \citealt{Jin.2017a, Jin.2021}).

We first perform this analysis for XMM-2 as its exposure time is much longer. We choose 0.3-1 keV as the reference band, and calculate the time lag and coherence in different frequency bins for 1-2, 2-4 and 4-10 keV, separately. The prescriptions in \citet{Vaughan.1997} and \citet{Nowak.1999} are followed to calculate the time lag, raw and Poisson-noise corrected coherences. We use the same gap-filling method described in Section~\ref{sec-rms} to minimize the influence of excluding background flares. We produce 1000 light curves, and calculate the average values to ensure the robustness of the final results. We emphasize that it is important to check the coherence whenever a phase/time lag is reported, because the lag is statistically meaningful only if there is a strong correlation between the two light curves.

Figure~\ref{fig-cohlag} shows the time lag and raw coherence in different frequency bins for the three energy bands. High coherences are found below $10^{-4}$ Hz, with significant positive lags found only in 4-10 keV. It shows that the variation in 4-10 keV lags behind 0.3-1 keV by $3.86\pm1.17$ ks in the lowest frequency bin, with the raw coherence being $0.70\pm0.15$. This is somewhat different from X-ray {\it simple} super-Eddington NLS1s, where a low-frequency hard X-ray lag is often found below 1 keV and increases towards softer X-rays (e.g. \citealt{Alston.2014, Jin.2017a, Jin.2020}).

Then we perform similar analysis for XMM-1, but in this case only the HF band can be covered, and the results are shown in Figure~\ref{fig-cohlag} in blue. The main difference in XMM-1 is the much higher HF coherence than in XMM-2, while the dependences of lag vs. frequency are similar. The lowest-frequency bin in XMM-1 shows a tentative lag of $0.75\pm0.31$ ks with a coherence of $0.53\pm0.22$, but this needs to be confirmed by future observations.

Finally, we focus on the LF band of $[0.95,5]\times10^{-5}$ Hz and explore the energy dependence in more detail. The 0.3-1 keV band is still chosen to be the reference band. Figure~\ref{fig-lagspec2} shows the time lag and coherence spectra. We find that the coherence is close to unity across the entire 0.3-10 keV band, indicating that the LF variability is strongly correlated between the soft and hard X-rays. The lag spectrum shows how the time lag depends on photon energy with a higher energy resolution. Again, only the energy band above 4 keV shows a significant lag. This is qualitatively consistent with a model where the energy band above 4 keV is dominate by a separate component, such as a neutral reflection component or a partial covering component (Figure~\ref{fig-specfit1}). We explore these models in more detail by fitting all the variability spectra simultaneously.

\begin{table}
 \centering
   \caption{Best-fit parameters of the neutral reflection plus Comptonisation model as shown in Figure~\ref{fig-specfit4} Panels a1 and a2. The total model is {\tt TBabs*zTBabs(compTT+nthComp+kdblur*pexmon)}. The total $\chi^2$ is 854.2 for 923 dof. {\it fixed} indicates that the parameter is fixed at the given value during the spectral fitting. {\it u} and {\it l} indicate that the parameter reaches its upper/lower limit. Errors are for the 90\% confidence limits.}
\begin{tabular}{lllr}
\hline
Component & Parameter & Value & Unit \\
\hline
 \multicolumn{4}{c}{\it XMM-2 Time-average Spectra} \\
 {\tt tbabs}     & $N_{\rm H}$ & 1.77 {\it fixed} & 10$^{20}$ cm$^{-2}$ \\
 {\tt ztbabs}    & $N_{\rm H}$ & 0.00 $^{+4.80}_{l}$  & 10$^{20}$ cm$^{-2}$ \\
 {\tt comptt}    & $kT_{\rm 0,s}$   &  8.93 $^{+1.17}_{-1.63}$  & $10^{1}$ eV \\
 {\tt comptt}   & $kT_{\rm e,s}$   &  0.25 $^{+0.04}_{-0.03}$ & keV \\
 {\tt comptt}    & $\tau$               & 1.21 $^{+0.14}_{-0.19}$ & $10^{1}$ \\
 {\tt comptt}    & norm        & 4.63 $^{+5.66}_{-0.80}$ & $10^{-3}$ \\
 {\tt nthcomp}   & $\Gamma$              & 2.27 $^{+0.06}_{-0.05}$ & \\
 {\tt nthcomp}    & $kT_{\rm0,h}$    & $=~kT_{\rm in}$ & eV \\
 {\tt nthcomp}   & $kT_{\rm e,h}$    & 200 {\it fixed} & keV \\
 {\tt nthcomp}   & norm      & 1.04 $^{+0.10}_{-0.12}$  & $10^{-4}$ \\
 {\tt kdblur}    & Index        & 3 {\it fixed}  & \\
 {\tt kdblur}    & $R_{\rm in}$         & 6 {\it fixed}  & $R_{\rm g}$ \\
 {\tt kdblur}    & $R_{\rm out}$       & 100 {\it fixed}  & $R_{\rm g}$ \\
 {\tt kdblur}   & $\theta_{\rm inc,1}$         & 60 {\it fixed}  & degree \\
 {\tt pexmon}   & $f_{\rm refl}$                   & -5.68 $^{+1.52}_{-2.11}$ & \\
 {\tt pexmon}   & $A_{\rm iron}$        & 1.00 $^{+0.77}_{-0.55}$ & \\
 {\tt pexmon}   & $\theta_{\rm inc,2}$   & $=~\theta_{\rm inc,1}$  & degree \\
 \multicolumn{1}{l}{$\chi^{2}$} & \multicolumn{3}{c}{459.7 for 515 bins} \\
 
  \multicolumn{4}{c}{\it XMM-2 HF-rms Spectrum} \\
  {\tt comptt}    & norm        & 6.27 $^{+7.90}_{-2.15}$ & $10^{-4}$ \\
  {\tt nthcomp}   & norm      & 1.09 $^{+0.36}_{-0.36}$  & $10^{-5}$ \\
 \multicolumn{1}{l}{$\chi^{2}$} & \multicolumn{3}{c}{15.0 for 8 bins} \\
   
  \multicolumn{4}{c}{\it XMM-2 LF-rms Spectrum} \\
  {\tt comptt}     & norm        & 9.64 $^{+11.4}_{-2.02}$ & $10^{-4}$ \\
  {\tt nthcomp}   & norm      & 1.34 $^{+0.25}_{-0.26}$  & $10^{-5}$ \\
 \multicolumn{1}{l}{$\chi^{2}$} & \multicolumn{3}{c}{8.3 for 8 bins} \\
   
  \multicolumn{4}{c}{\it XMM-1 Time-average Spectrum} \\
  {\tt comptt}    & norm        & 1.41 $^{+1.87}_{-0.20}$ & $10^{-2}$ \\
  {\tt nthcomp}   & norm      & 5.79 $^{+0.37}_{-0.49}$  & $10^{-4}$ \\
  {\tt pexmon}   & rel\_refl       & -0.00 $^{u}_{-1.51}$ & \\
  \multicolumn{1}{l}{$\chi^{2}$} & \multicolumn{3}{c}{360.8 for 402 bins} \\
  
  \multicolumn{4}{c}{\it XMM-1 HF-rms Spectrum} \\
  {\tt comptt}    & norm        & 8.06 $^{+11.6}_{-8.06}$ & $10^{-4}$ \\
  {\tt nthcomp} & norm      & 1.42 $^{+0.11}_{-0.12}$  & $10^{-4}$ \\
 \multicolumn{1}{l}{$\chi^{2}$} & \multicolumn{3}{c}{10.5 for 8 bins} \\
\hline
   \end{tabular}
 \label{tab-xrayfit1}
\end{table}

\begin{table}
 \centering
   \caption{Best-fit parameters of the partial covering absorption model as shown in Figure~\ref{fig-specfit4} Panels b1 and b2. The total model is {\tt zTBabs*TBabs(powerlaw1+zTBabs2*powerlaw2)}. {\tt powerlaw1} is the primary component, and {\tt powerlaw2} is the more severely absorbed component. The total $\chi^2$ is 876.3 for 927 dof. {\it fixed} indicates that the parameter is fixed during the spectral fitting. {\it l} indicates that the parameter reaches its lower limit. Errors are for the 90\% confidence limits. }
\begin{tabular}{lllr}
\hline
Component & Parameter & Value & Unit \\  
\hline
 \multicolumn{4}{c}{\it XMM-2 Time-average Spectra} \\
 {\tt TBabs}     & $N_{\rm H}$ & 1.77 {\it fixed} & 10$^{20}$ cm$^{-2}$ \\
 {\tt ztbabs}    & $N_{\rm H}$ & 0.00 $^{+0.55}_{l}$  & 10$^{20}$ cm$^{-2}$ \\
 {\tt powerlaw1}   & $\Gamma_1$              & 2.24 $^{+0.03}_{-0.02}$ & \\
 {\tt powerlaw1}   & norm              & 1.29 $^{+0.02}_{-0.02}$ & $10^{-4}$ \\
 {\tt zTBabs2}   & $N_{\rm H}$ & 3.25 $^{+1.20}_{-0.95}$  & 10$^{23}$ cm$^{-2}$ \\
 {\tt powerlaw2}   & $\Gamma_2$              & $=~\Gamma_1$ & \\
 {\tt powerlaw2}   & norm              & 1.49 $^{+0.30}_{-0.26}$ & $10^{-4}$ \\
 \multicolumn{1}{l}{$\chi^{2}$} & \multicolumn{3}{c}{469.0 for 515 bins} \\
 
  \multicolumn{4}{c}{\it XMM-2 HF-rms Spectrum} \\
 {\tt powerlaw1}  & norm              & 1.44 $^{+0.22}_{-0.22}$ & $10^{-5}$ \\
 {\tt powerlaw2}  & norm              & 2.51 $^{+2.31}_{-2.16}$ & $10^{-5}$ \\
 \multicolumn{1}{l}{$\chi^{2}$} & \multicolumn{3}{c}{12.2 for 8 bins} \\
 
   \multicolumn{4}{c}{\it XMM-2 LF-rms Spectrum} \\
 {\tt powerlaw1}  & norm              & 2.00 $^{+0.13}_{-0.13}$ & $10^{-5}$ \\
 {\tt powerlaw2}  & norm              & 0.00 $^{+0.46}_{l}$ & $10^{-5}$ \\
 \multicolumn{1}{l}{$\chi^{2}$} & \multicolumn{3}{c}{20.2 for 8 bins} \\
 
 \multicolumn{4}{c}{\it XMM-1 Time-average Spectra} \\
 {\tt powerlaw1}   & $\Gamma_3$              & $=~\Gamma_1$ & \\
 {\tt powerlaw1}   & norm              & 4.35 $^{+0.32}_{-0.40}$ & $10^{-4}$ \\
 {\tt zTBabs2}   & $N_{\rm H}$ & 3.13 $^{+1.05}_{-0.92}$  & 10$^{21}$ cm$^{-2}$ \\
 {\tt powerlaw2}   & $\Gamma_4$              & $=~\Gamma_1$ & \\
 {\tt powerlaw2}   & norm              & 2.85 $^{+0.37}_{-0.35}$ & $10^{-4}$ \\
 \multicolumn{1}{l}{$\chi^{2}$} & \multicolumn{3}{c}{361.1 for 402 bins} \\
 
   \multicolumn{4}{c}{\it XMM-1 HF-rms Spectrum} \\
 {\tt powerlaw1}  & norm              & 4.50 $^{+1.71}_{-2.34}$ & $10^{-5}$ \\
 {\tt powerlaw2}  & norm              & 1.36 $^{+0.25}_{-0.24}$ & $10^{-4}$ \\
 \multicolumn{1}{l}{$\chi^{2}$} & \multicolumn{3}{c}{13.8 for 8 bins} \\
\hline
   \end{tabular}
 \label{tab-xrayfit2}
\end{table}

\section{Modelling the Full X-ray Spectral Variability}
\label{sec-xvarfit0}
\subsection{Modelling the Time-average and Rms Spectra}
\label{sec-xvarfit1}
To maximized the model constraints, we perform a simultaneous modelling of the time-average spectra and rms spectra of both flux states. We also assume that the spectral components in XMM-1 and XMM-2 have the same shape, only that their normalizations are different. The physical assumption is that we see different fractions of each component in different observations. Then we can fit the five spectral simultaneously, including the time-average spectrum and the HF rms spectrum of XMM-1, the time-average spectra (including both \xmm\ and \nustar\ spectra), LF and HF rms spectra of XMM-2. Considering the shapes pf the LF rms spectrum and lag spectrum in XMM-2, it is obvious that the soft and hard X-rays cannot be dominated by the same component. This means that the ionized reflection model can be clearly ruled out, because it produces significant fluxes in both soft and hard X-ray bands (see Figure~\ref{fig-xrayspec}a). Therefore, we only consider the neutral reflection plus Comptonisation model and the partial covering absorption model. 

\begin{figure*}
\centering
\begin{tabular}{cc}
\includegraphics[trim=-0.1in 0.1in 0.0in 0.0in, clip=1, scale=0.48]{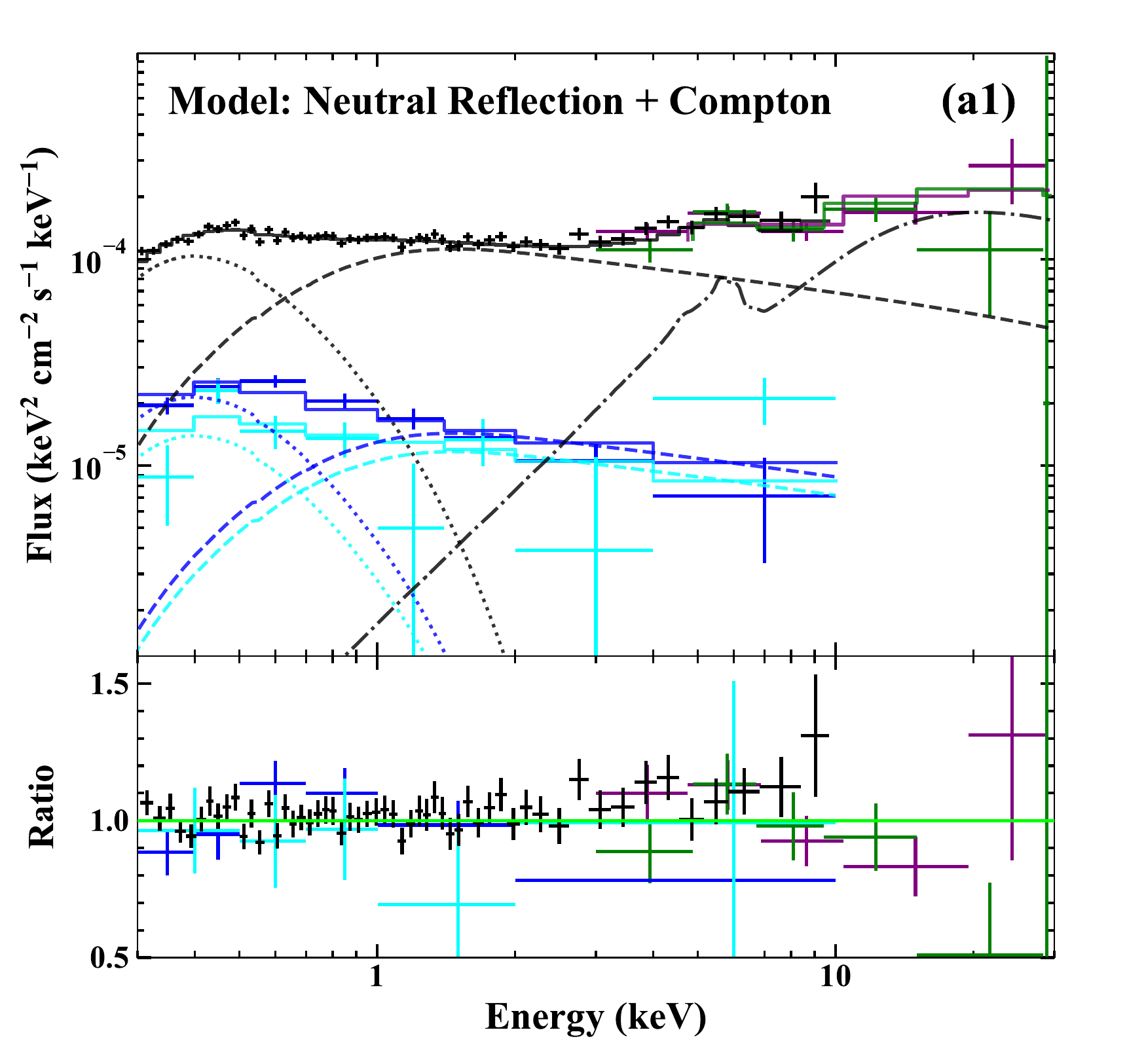} &
\includegraphics[trim=0.4in 0.1in 0.0in 0.0in, clip=1, scale=0.48]{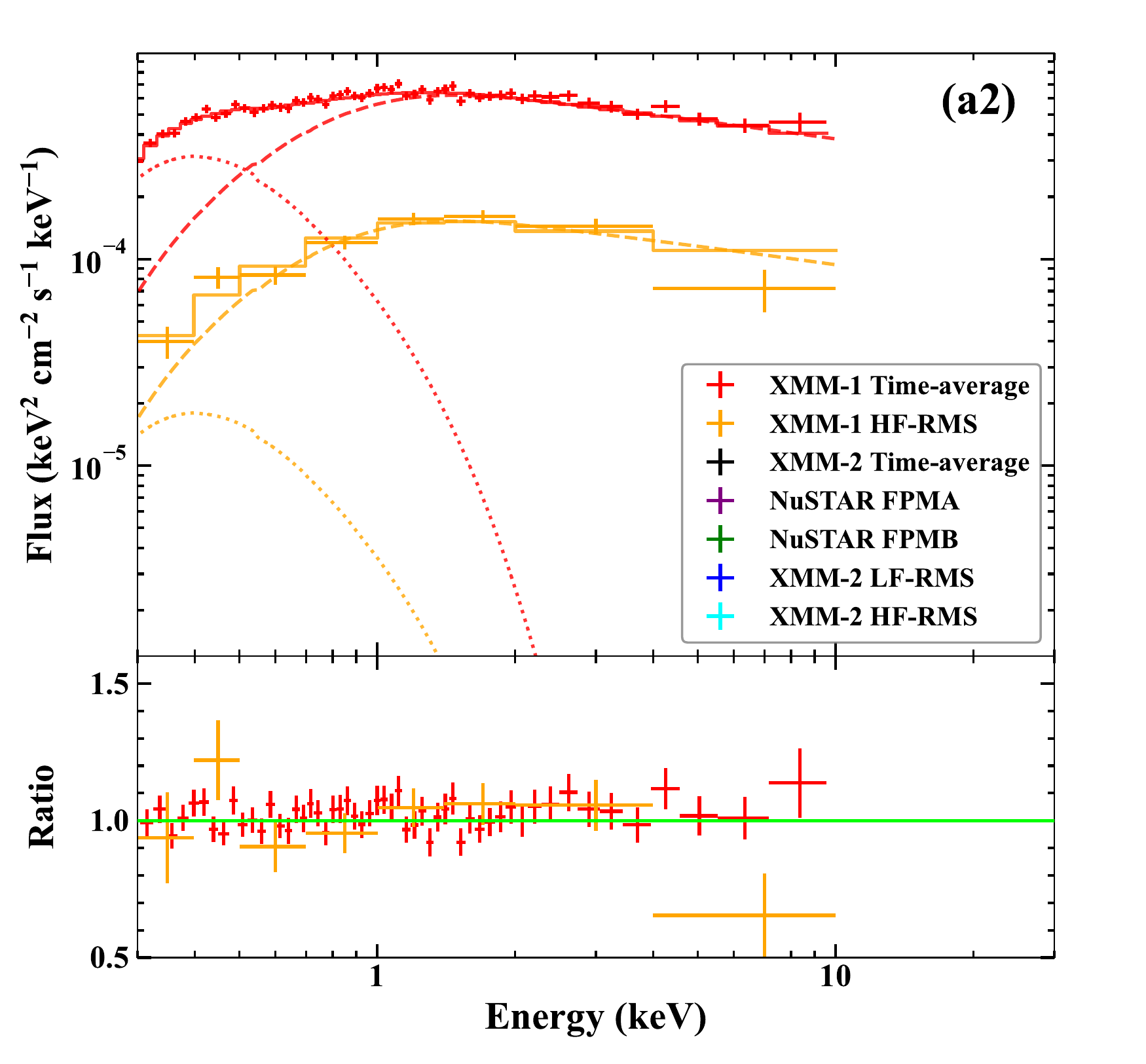} \\
\includegraphics[trim=-0.1in 0.1in 0.0in 0.0in, clip=1, scale=0.48]{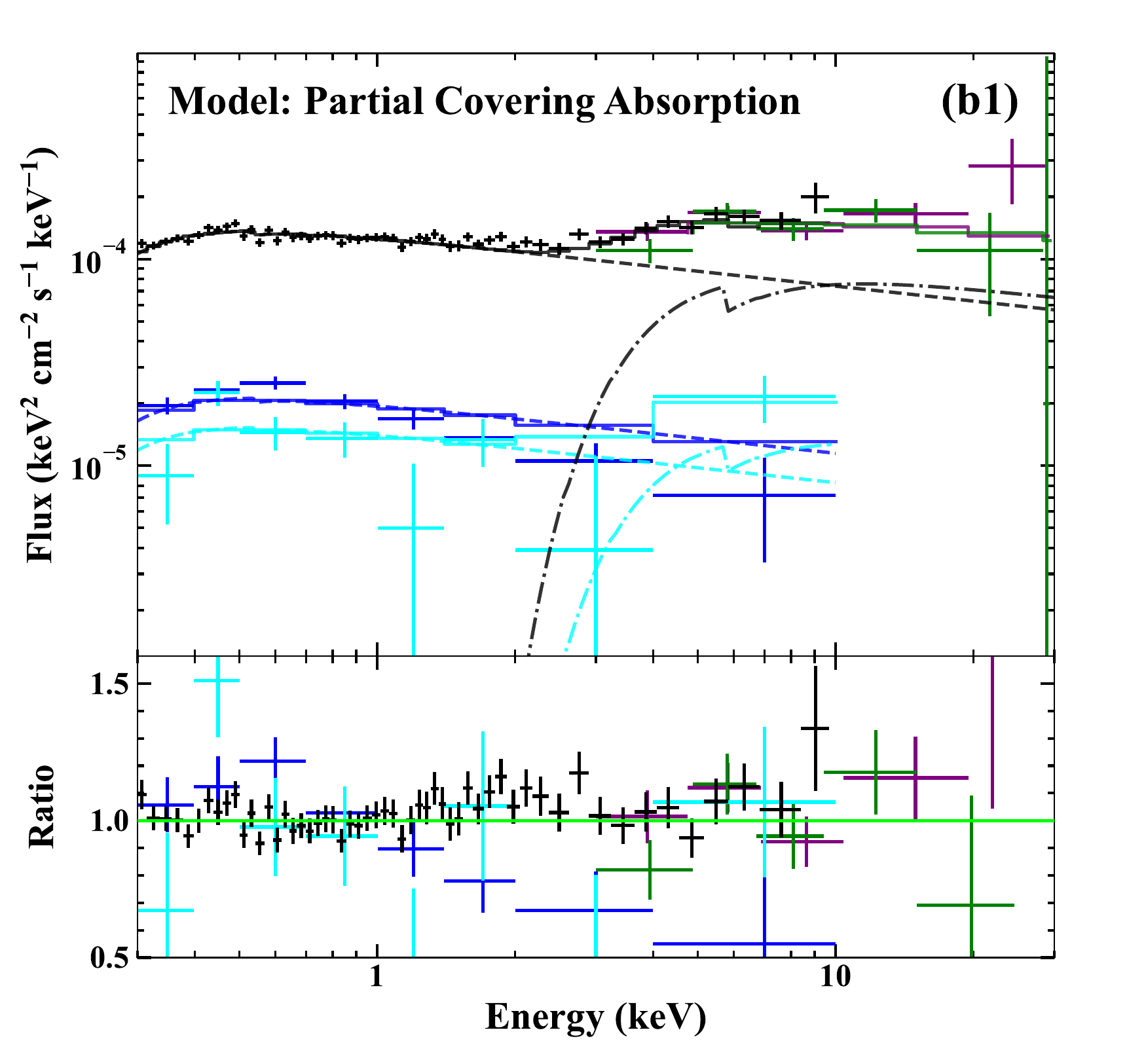} &
\includegraphics[trim=0.4in 0.1in 0.0in 0.0in, clip=1, scale=0.48]{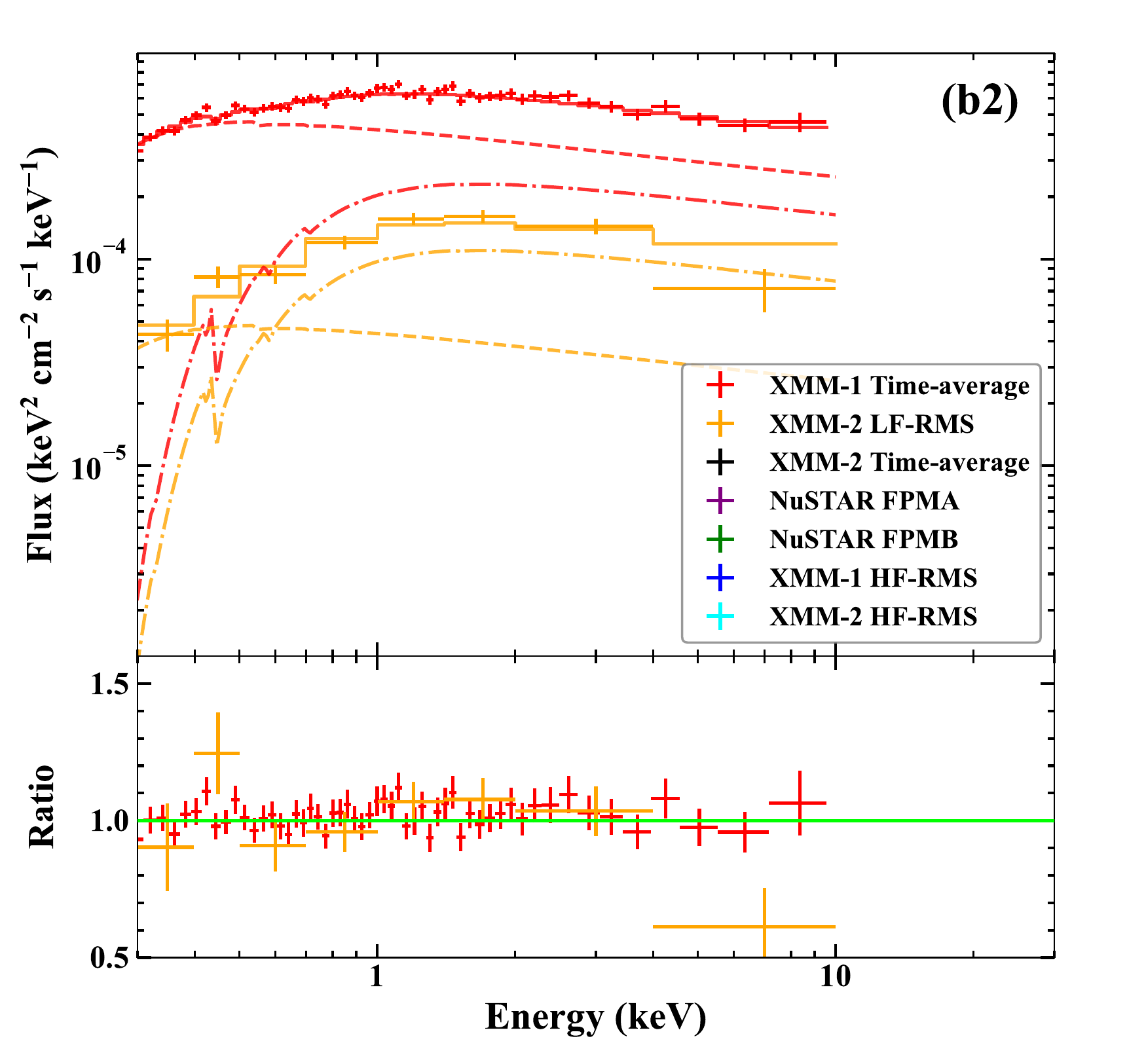} \\
\end{tabular}
\caption{Simultaneous fits to all the spectra of \rxj0134, including the time-average spectra in XMM-1 (red) and XMM-2 (black), LF rms (blue) and HF rms (cyan) spectra in XMM-2, HF rms spectrum in XMM-1 (orange). Panels a1 and a2 are for the neutral reflection plus Comptonisation model, including a warm Comptonisation (dotted line), a hot Comptonisation (dash line) and a neutral reflection (dash-dot line). These components have the same shape but different normalizations in different spectra. Panels b1 and b2 are for the partial covering absorption model, including a primary power law (dash line) and a more severely absorbed power law (dash-dot line).}
\label{fig-specfit4}
\end{figure*}

For the neutral reflection plus Comptonisation model, we obtain a total $\chi^2$ of 854.2 for 923 dof for all the five spectra, indicating that this model can fit all the spectra simultaneously. As shown in Figure~\ref{fig-specfit4}, during the low-flux state of XMM-2, the reflection component contributes significantly to the time-average \xmm\ and \nustar\ spectra above 4 keV. The reflection fraction parameter is found to be $f_{\rm refl} = -5.68^{+1.52}_{-2.11}$, which suggests that the reflection materials may see a larger fraction of the hot corona than us. But during the high-flux state of XMM-1, the reflection component is not detected, although the quality of the XMM-1 spectrum only allows us to rule out the same reflection flux as in XMM-2 with 2.0 $\sigma$ significance. These results imply that the low-flux state of XMM-2 might be caused by an episodic obscuration of the coronal emission along the line of sight.

In the time-average spectra of both XMM-1 and XMM-2, there is a warm Comptonisation component in the soft X-rays, whose electron temperature is $0.25^{+0.04}_{-0.03}$ keV and optical depth is $12.1^{+1.4}_{-1.9}$. The hard X-ray continuum can be well fitted by a hot Comptonisation component with a photon index of $\Gamma=2.27^{+0.06}_{-0.05}$. These parameters are all similar to typical values of X-ray {\it simple} super-Eddington NLS1s, as shown in Table~\ref{tab-xrayfit1}. All the rms spectra can also be well fitted by the two Comptonisation components. Similarly, this implies that \rxj0134\ could intrinsically be a typical X-ray {\it simple} super-Eddington NLS1 with a steep hard X-ray slope and a typical soft excess. But in this case we only observe a small fraction of its soft excess, which is probably also caused by the obscuration of the coronal emission.

The partial covering absorption model produces comparably good fits to all the spectra, with a total $\chi^2$ of 876.3 for 927 dof. As shown in Figure~\ref{fig-specfit4}, during the high-flux state of XMM-1, the time-average spectrum is well fitted by the two power law components. The primary power law has a photon index of $\Gamma=2.24^{+0.03}_{-0.02}$ with no intrinsic absorption. This power law dominates the entire 0.3-10 keV band, although an increasing contribution from the secondary power law is observed towards hard X-rays, and it is absorbed by an additional column density of $N_{\rm H}=3.13^{+1.05}_{-0.92}\times10^{21}$ cm$^{-2}$. During the low-flux state of XMM-2, the time-average spectra below 4 keV is totally dominated by the primary power law with the same shape, while the secondary power law is absorbed severely by an a much higher column density of $N_{\rm H} = 3.25^{+1.20}_{-0.95}\times10^{23}$ cm$^{-2}$, and so it mainly contributes to the spectra above 4 keV.

In this partial covering absorption model, the LF and HF rms spectra of XMM-2 are both dominated by the primary power law. But the fit to the LF rms spectrum has $\chi^2=20.2$, which is significantly larger than $\chi^2=8.3$ as we found in the neutral reflection plus Comptonisation model. This is also visible in Figure~\ref{fig-specfit4}b1, where the LF rms spectrum (in blue) is more curved than the primary power law. The goodnesses of fits to the other two rms spectra are comparable between the two models (see Table~\ref{tab-xrayfit2}). 

We also note that in XMM-2 the primary power law has a normalization of $1.29\times10^{-4}$ in the time-average spectrum and $1.44\times10^{-5}$ in the HF rms spectrum, which indicates that this power law has 11\% flux variation at HF in XMM-2. Interestingly, in XMM-1 this primary power law has a normalization of $4.35\times10^{-4}$ in the time-average spectrum and $4.50\times10^{-5}$ in the HF rms spectrum, which indicates a flux variation of 10 per cent at HF in XMM-1, which is fully consistent with XMM-1. However, the HF rms spectrum of XMM-1 is mainly dominated by the secondary power law, suggesting that this component is more variable than the primary power law. This can be understood if the materials causing stronger obscuration of the secondary power law also introduces more variability at HF, although the exact mechanism for this is not clear.

\subsection{Modelling the X-ray Lag Spectrum}
\label{sec-xvarfit2}
With the above two best-fit models, we can now test whether they reproduce the lag spectrum in Figure~\ref{fig-lagspec2}. This can be done by fitting the lag spectrum by introducing phase lags between different spectral components at specific frequency bands. This approach was first introduced by \citet{Jin.2021} to model the lag spectrum of \rej1034. Here we apply it to the two best-fit models of \rxj0134.

As shown in Table~\ref{tab-timelag}, the neutral reflection plus Comptonisation model contains three components, including {\tt compTT}, {\tt nthcompt} and {\tt pexmon}. We use the observed light curves in 0.3-0.6 keV and 1-2 keV to represent the intrinsic variability of the two Comptonisation components. The incident emission for {\tt pexmon} is assumed to be {\tt nthcompt}, so we also use the light curve in 1-2 keV to represent its intrinsic variability. Then We can modify the phase of these light curves in the Fourier domain in $[0.95,5]\times10^{-5}$ Hz to produce different lags. Then the total lag in every energy bin can be calculated using the relative contribution of the three components in the best-fit model. The merit of this method is that it avoids the uncertainty from creating artificial light curves.

The blue dash line in Figure~\ref{fig-lagspec2}a shows the best-fit lag model in the neutral reflection scenario, with a minimal $\chi^2$ of 6.8 for 7 dof (see Table~\ref{tab-timelag}). The significant hard lag can be well reproduced by the lag of the neutral reflection component. The required intrinsic lag is 7.54$\pm$2.91 ks between {\tt pexmon} and {\tt compTT}. This large intrinsic time lag is also consistent with the reflection being located at a distant region with a low ionization state. Additionally, {\tt nthcompt} is found to lag behind {\tt compTT} by 1.23$\pm$0.72 ks, which is consistent with a propagation lag, similar to the observation of normal super-Eddington NLS1s such as \rx04\ (\citealt{Jin.2017a}).

The partial covering absorption model contains two components, including the primary power law and a more absorbed secondary power law. We use the light curve in 0.3-2 keV to represent the intrinsic light curve of the primary power law. Since the secondary power law is considered to have the same origin as the primary power law, we also use the same 0.3-2 keV light curve to represent its intrinsic variability. After running a fit to the lag spectrum, the minimal $\chi^2$ is found to be 10.6 for 8 dof, which is slightly worse than the neutral reflection model by 1.9$\sigma$ significance. As shown by the red dash line in Figure~\ref{fig-lagspec2}a, the significant hard lag above 4 keV can also be well reproduced, with the secondary power law lagging behind the primary power law by 8.81$\pm$3.14 ks. The main difference from the neutral reflection scenario is that no lag can be produced below 2 keV in this absorption scenario, which is because the spectrum below 2 keV is completely dominated by the primary power law alone.

A possible mechanism of the time lag in the partial covering absorption scenario is the interaction between the intrinsic emission and the disc wind, which is similar to the wind-driven mechanism proposed to explain the spectral variability of NGC 1365 (\citealt{Connolly.2014}). In this picture, there is a light travel time (i.e. the lag time) between the source of intrinsic emission and the disc wind. If the intrinsic emission enhances, the wind will respond to it with a time lag by decreasing its column density, and then the absorption component will also be enhanced. 

However, it is more natural to explain the lag within the neutral reflection scenario. This model also produces slightly better fit to the LF lag spectrum than the partial covering absorption model, although the improvement of $\chi^2$ only has 1.9$\sigma$ significance. As shown in Figure~\ref{fig-lagspec2}a, the 4 ks lag above 4 keV can be reproduced by both models, but the additional warm Comptonisation component in the neutral reflection model produces slightly better fit to the lag below 2 keV. Considering that the warm Comptonisation model is also preferred by normal X-ray {\it simple} super-Eddington NLS1s (e.g. \citealt{Jin.2009, Jin.2013, Jin.2016, Jin.2017a, Kara.2017, Parker.2018}), the intrinsic properties of \rxj0134\ may not be very different from the other super-Eddington NLS1s, except that we may be viewing it along a special line of sight to its intrinsic emission. We will discuss this in more detail in the next section.

\begin{table}
 \centering
   \caption{Best-fit time lags from the lag spectrum modelling in Figure~\ref{fig-lagspec2} with different spectral models. {\it ref} indicates the component is treated as the reference component. A positive time lag indicates the component lags behind the reference component.}
\begin{tabular}{@{}lcccc@{}}
\hline
& \multicolumn{4}{c}{\it Model: Neutral Reflection + Comptonisation}\\
Component & {\tt comptt} & {\tt nthcomp} & {\tt pexmon} & $\chi^{2}_{\nu}$ \\
Time Lag & {\it ref} & 1.23 $\pm$ 0.72 ks & 7.54 $\pm$ 2.91  ks & 6.8/7 \\
\hline
& \multicolumn{4}{c}{\it Model: Partial Covering Absorption}\\
Component & {\tt powerlaw1} & \multicolumn{2}{c}{\tt powerlaw2} & $\chi^{2}_{\nu}$ \\
Time Lag & {\it ref} & \multicolumn{2}{c}{8.81 $\pm$ 3.14 ks} & 10.6/8 \\
\hline
   \end{tabular}
 \label{tab-timelag}
\end{table}

\begin{table}
\centering
   \caption{Black hole mass estimates of \rxj0134\ using the $\sigma^{2}_{\rm rms}$-$M_{\rm BH}$ relations for different timescales in \citet{Ponti.2012}. The infinity value indicates that the upper/lower limit is unconstrained. Segment `all' means that the entire light curve is used. Different rms can be measured from different frequency ranges corresponding to different timescales.}
     \begin{tabular}{lccc}
     \hline
    Observation & Segment & Rms & $M_{\rm BH}$ \\
     & ({\it ksec}) & & ($10^{7}M_{\odot}$) \\
    \hline
    \multicolumn{4}{l}{\it Timescale: 20 ks}\\
    XMM-1 & 0-20 & 0.232 $^{+0.018}_{-0.018}$ & 0.20 $^{+0.03}_{-0.02}$\\
    XMM-2 & 0-20 & 0.058 $^{+0.065}_{-0.058}$  & 1.89 $^{+\infty}_{-1.33}$ \\
    XMM-2 & 20-40 & 0.072 $^{+0.054}_{-0.054}$ & 1.33 $^{+11.2}_{-0.79}$ \\
    XMM-2 & 40-60 & 0.097 $^{+0.048}_{-0.048}$ & 0.83 $^{+1.66}_{-0.39}$ \\
    XMM-2 & 60-80 & 0.063 $^{+0.064}_{-0.063}$ &1.66 $^{+\infty}_{-1.12}$ \\
    XMM-2 & 80-100 & 0.075 $^{+0.058}_{-0.058}$ & 1.25 $^{+12.4}_{-0.75}$ \\
    XMM-2 & all & 0.060 $^{+0.032}_{-0.032}$ & 1.80 $^{+4.50}_{-0.90}$ \\
    \hline
    \multicolumn{4}{l}{\it Timescale: 40 ks}\\
    XMM-2 & 10-50 & 0.054 $^{+0.047}_{-0.047}$ & 2.54 $^{+54.7}_{-1.77}$ \\
    XMM-2 & 60-100 & 0.068 $^{+0.047}_{-0.047}$ & 1.79 $^{+7.94}_{-1.17}$ \\
    XMM-2 & all & 0.060 $^{+0.032}_{-0.032}$ & 2.17 $^{+3.95}_{-1.27}$ \\
    \hline
    \multicolumn{4}{l}{\it Timescale: 80 ks}\\
    XMM-2 & 20-100 & 0.105 $^{+0.026}_{-0.026}$ & 1.04 $^{+0.11}_{-0.53}$ \\
    XMM-2 & all & 0.069 $^{+0.030}_{-0.030}$ & 2.15 $^{+1.44}_{-1.35}$ \\
    \hline
     \end{tabular}
\label{tab-rms-mass}
\end{table}

\section{Discussion}
\subsection{Comparison of X-ray Spectra with Typical NLS1s}
\label{sec-discussion1}
\rxj0134\ is a typical extreme super-Eddington NLS1 in terms of its black hole mass and Eddington ratio. We find that its X-ray time-average spectrum in the high-flux state can be well fitted by a steep power law with $\Gamma \simeq 2.2$ (see Section~\ref{sec-xray-spec}). The spectrum in the low-flux state of XMM-2 remains steep, with a very weak soft excess below 0.6 keV and a strong hard excess above 4 keV (see Figure~\ref{fig-xrayspec}). The smoothness of the spectra is similar to those observed in X-ray {\it simple} NLS1s. Below we compare the X-ray spectral shape between \rxj0134\ and three X-ray {\it simple} NLS1s.

\rx04\ is an archetypal super-Eddington NLS1 with a black hole mass of $M_{\rm BH}=7\times10^{6}M_{\odot}$ and an Eddington ratio of $L_{\rm bol}/L_{\rm Edd}=4.6$ (\citealt{Jin.2017a, Jin.2017b}). \rej1034\ is a super-Eddington NLS1 with one of the strongest soft excesses and has $M=2\times10^{6}M_{\odot}$ and $L_{\rm bol}/L_{\rm Edd}=2.2$ (\citealt{Jin.2021}). It also shows the most robust X-ray quasi-periodic oscillation (QPO) signal in AGN (\citealt{Gierlinski.2008, Middleton.2011, Alston.2014, Jin.2020}). \pg12\ is also a typical X-ray {\it simple} super-Eddington NLS1 with $M_{\rm BH}=5\times10^{6}M_{\odot}$ and $L_{\rm bol}/L_{\rm Edd}=2.6$, although it is not as extreme as \rx04\ in terms of its Eddington ratio and soft excess (\citealt{Jin.2013, Done.2016}). These AGN are representative of the X-ray {\it simple} super-Eddington NLS1 population.

For the purpose of spectral comparison, we apply the Comptonisation model for \rxj0134\ and the other three NLS1s to produce the unfolded spectra. \rxj0134\ exhibits a significant excess above 4 keV in XMM-2, thus we add a neutral reflection component to improve the fit. This component is not included in the fits of the other sources, because their fits are already statistically plausible. However, it does not mean that Comptonisation plus reflection must be the correct model for all the sources. Indeed, different models have been proposed for these sources based on more detailed X-ray spectral-timing studies on a case by case basis (e.g. \citealt{Jin.2013, Kara.2014, Jin.2017a, Jin.2021}), but it does not affect our comparison of the spectral shapes. Then we correct these spectra for their individual Galactic extinction, and then de-redshift them to their respective rest-frames. Then we rescale all the spectra to the flux of \rxj0134\ at 3 keV in XMM-1, so that we can directly compare their soft and hard X-ray shapes.

\begin{figure}
\centering
\includegraphics[trim=0.05in 0.3in 0.0in -0.1in, clip=1, scale=0.57]{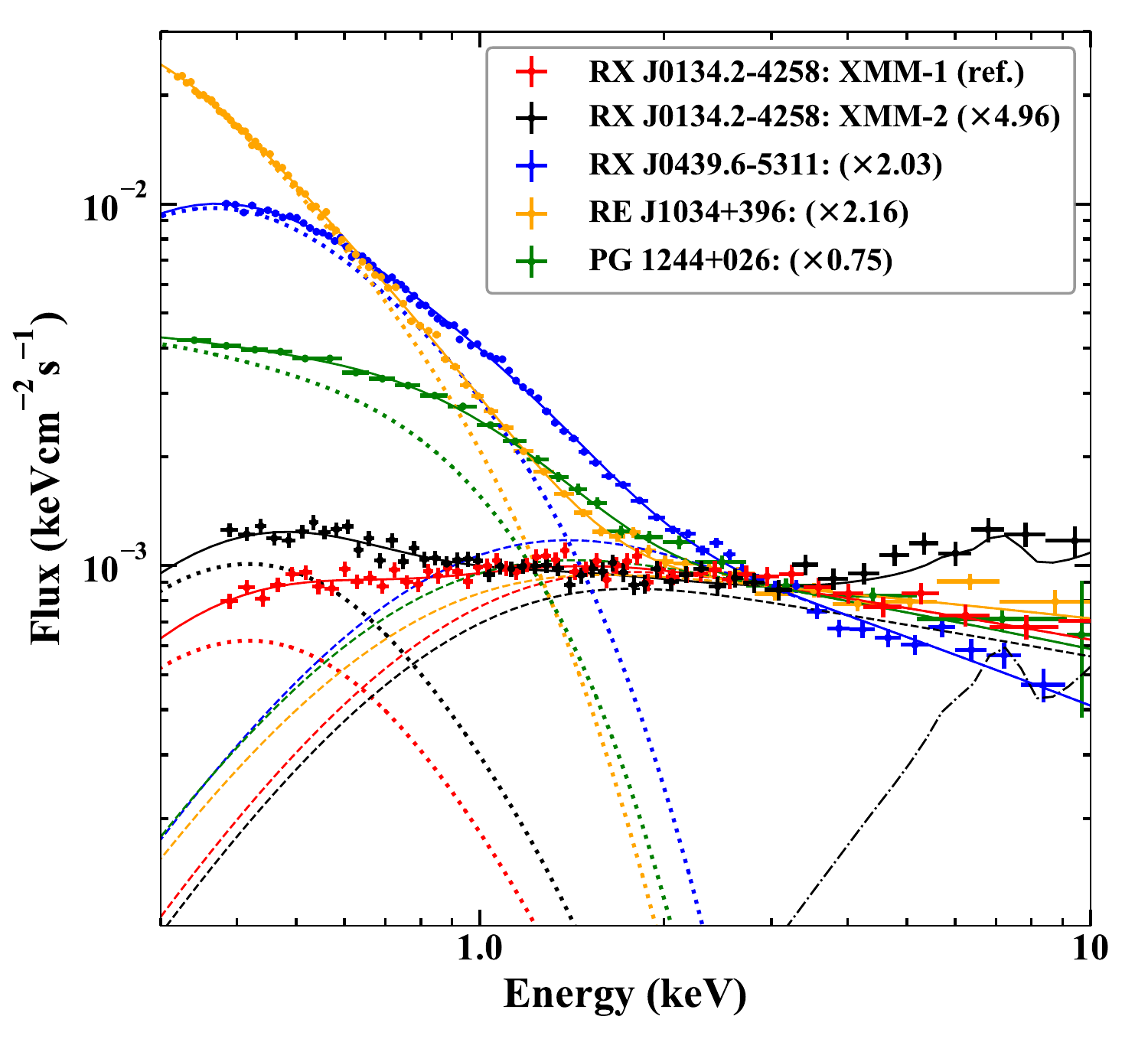} 
\caption{Comparison of the X-ray unfolded spectra of \rxj0134\ and some representative X-ray {\it simple} super-Eddington NLS1s. All the spectra are fitted with a worm Comptonisation component (dotted line) in the soft X-rays and a hot Comptonisation component (dash line) in the hard X-rays. An additional neutral reflection component (dash-dot line) is included for the XMM-2 spectrum of \rxj0134\ to improve the spectral fit above 4 keV. All the spectra have been corrected for their respective Galactic extinctions, de-redshifted to their rest-frames, and renormalized to the XMM-1 spectrum (red) of \rxj0134\ at 3 keV. The applied scaling factors are given in the legend.}
\label{fig-sx-compare}
\end{figure}

As shown in Figure~\ref{fig-sx-compare}, all the spectra appear smooth and can be well fitted by the two Comptonisation components. The shape of the hot Comptonisation is similar in all the sources, including a steep slope and a low-energy cut-off. Since we assume that the warm corona provides seed photons for the hot corona, the similar cut-off energy indicates a similar warm corona temperature in all the sources, which is in the range of 0.2-0.25 keV (\citealt{Czerny.2003, Gierlinski.2004, Petrucci.2018}). The shape of the warm Compensation of \rxj0134\ is very similar to \rx04, but its intensity relative to the hard X-ray emission is about one order of magnitude lower. The excess flux above 4 keV as seen in the low-flux state of \rxj0134\ in XMM-2 is also remarkable. X-ray {\it simple} super-Eddington NLS1s generally show a steep hard X-ray slope with weak or non-detectable reflection features (e.g. an iron K$\alpha$ line). Thus the non-detection of this component in the high-flux state of XMM-1 is more consistent with typical X-ray {\it simple} NLS1s.

While the smooth spectra of \rxj0134\ resemble X-ray {\it simple} NLS1s, its drastic X-ray variability is more similar to some X-ray {\it complex} super-Eddington NLS1s, such as \1h0707\ and IRAS 13224-3809, whose X-ray emission show drastic variation in terms of both flux and spectral shape \citealt{Done.2016, Jiang.2018, Parker.2021, Boller.2021}). Besides, extreme X-ray variability is also observed in some weak-line quasars (WLQs, \citealt{Miniutti.2012, Ni.2020}). Note that these AGN all have high Eddington ratios, thus their extreme X-ray variability may be caused by similar mechanisms, such as complex obscurers along the line of sight, which may be related to the clumpy disc wind and/or puffed-up inner accretion disc (e.g. \citealt{Luo.2015, Hagino.2016, Jin.2017b, Ni.2018}).

Moreover, the optical-to-X-ray spectral index ($\alpha_{\rm ox}$, e.g \citealt{Lusso.2010}) of \rxj0134\ is measured to be 1.4 and 1.7 in the high and low-flux states (see Paper-II), both of which are much smaller than $\alpha_{\rm ox}=2.3$ as observed in the archetypal X-ray weak WLQ PHL 1811 (e.g. \citealt{Leighly.2007}). This means that although the soft excess of \rxj0134\ is extremely weak compared to typical NLS1s, its X-ray emission is still much stronger than expected for regular X-ray weak AGN. Hence, \rxj0134\ may represent a new special type of AGN. We will present more detailed multi-wavelength comparison for these AGN in Paper-II.

\subsection{Peculiar X-ray Spectral States and Mechanisms}
\label{sec-mechanism}
In this work, we have proposed two possible scenarios to model the spectral-timing properties of \rxj0134\ as observed in the two distinct spectral states of XMM-1 and XMM-2. However, there remains some unclear issues in both scenarios.

In the neutral reflection plus Comptonisation scenario, the hot corona produces the primary hard X-ray emission, while the warm corona produces the soft excess. The interpretation for the hard X-ray excess is the reflection of intrinsic coronal emission by the wind materials at large radii. The distance of this reflection can be inferred from the time lag of 7.54 $\pm$ 2.91 ks to be 50-100 $R_{\rm g}$ for $M_{\rm BH}=2.0\times10^{7}M_{\odot}$. A possible explanation for the low-flux state of XMM-2 is the partial obscuration of the primary emission. But there is no sign of severe absorption in this scenario in the data (see Figure~\ref{fig-specfit4} Panels a1 and a2), so the obscuration materials should be Compton thick and have a clear boundary. Such materials could be provided by the clumpy disc wind, or the puffed-up disc itself for special lines of sight (e.g. \citealt{Luo.2015, Jin.2017b}). In this scenario, the change of covering fraction leads to the two distinct spectral states. Then an obvious prediction is that the reflection component does not change as much, because it comes from a more extended region. Another possible explanation is the intrinsic dimming of the primary emission from the high-flux state to the low-flux state. Then the intensity of the reflection component should also decrease. It must be noted that these two explanations are not mutually exclusive. Indeed, since the intensity of the primary emission and the covering fraction may be physically correlated, it is also possible that the change of flux state is caused by the variation of both the primary emission and the obscuration. Unfortunately, due to the low spectral quality of XMM-1, we cannot obtain a strong constraint on the potentially underlying reflection component in the high-flux state (see Section~\ref{sec-xvarfit1}). Thus we have to wait for deeper \xmm\ and \nustar observations in the high-flux state to verify this scenario in the future.

In the partial covering absorption scenario, the change of flux state is caused by the change of both column density and the covering fraction. As shown by the best-fit parameters in Table~\ref{tab-xrayfit2}, as \rxj0134\ changes from the high-flux state to the low-flux state, the column density of the absorber increases from $3.13\times10^{21} cm^{-2}$ to $3.25\times10^{23} cm^{-2}$. Likewise, the flux drop of the primary power law can be a result of enhanced obscuration and/or intrinsic dimming. If we assume that there are no other types of obscurers, then the covering fraction of the absorber, which is calculated from the ratio of normalization of the two power law components, increases from 40 to 56 per cent during this flux-state transition.

The {\it ASCA} spectrum appears to be an intermediate state between the low and high-flux states observed by \xmm. Since its data quality is low, it is easy to obtain good spectral fits with different model scenarios. In the first scenario, the rather hard spectral shape may suggest a weaker primary emission than XMM-1 but stronger reflection than XMM-2. While in the second scenario, the spectral shape may lead to a much larger covering fraction of the absorber. However, we prefer not to speculate too much about the origin of this intermediate state, because the current data cannot provide stronger constraints on the X-ray mechanism, unless better data of this flux state are available in the future.

Finally, we must emphasize that the X-ray luminosity of \rxj0134\ only contributes a tiny fraction of its bolometric luminosity. For example, the observed 0.3-10 keV luminosity is only $\sim$2 per cent of the bolometric luminosity. This is because the broadband spectral energy distribution of \rxj0134, which was presented in \citet{Grupe.2010}, exhibits a strong peak in the UV, which contains most of the bolometric luminosity. Thus different intrinsic X-ray luminosities inferred by different scenarios will not affect the bolometric luminosity much. \rxj0134\ remains a robust super-Eddington NLS1 regardless of its X-ray spectral states. This is similar to some X-ray {\it complex} super-Eddington NLS1s such as 1H 0707-495, whose X-ray emission shows drastic variability, but the simultaneous optical/UV emission is quite stable (\citealt{Done.2016}).

\subsection{Short-term X-ray Variability in XMM-1 and XMM-2}
\label{sec-short-xvar}
The short-term X-ray variability of \rxj0134\ appears very different between XMM-1 and XMM-2, but we cannot rule out the influence of severe red noise leak in XMM-1. Here we investigate this issue further by following the prescriptions in \citet{Vaughan.1997} and \citet{Arevalo.2008} to calculate the rms in each observation. Then we can use these rms values to get different black hole mass estimates, i.e. using the $\sigma^2_{\rm rms}$-$M_{\rm BH}$ relation (\citealt{Lu.2001, Zhou.2010, Ponti.2012, Jin.2016}). After that we can compare them with the masses estimated by other methods, thereby allowing the true rms to be inferred. The black hole mass of \rxj0134\ was estimated to be $M_{\rm BH}~=~1.47\times10^{7}M_{\odot}$, which is a virial mass based on the single-epoch H$\beta$ full width at half maximum (FWHM) of 1160 km s$^{-1}$, as well as the radius-luminosity (R-L) relation reported by \citet{Kaspi.2000}. This black hole mass sets a benchmark for comparison.

\citet{Ponti.2012} also showed that for different timescales, the coefficients of this $\sigma^2_{\rm rms}$-$M_{\rm BH}$ relation are different. Thus we also calculate the rms for different timescales of 20, 40 and 80 ks separately. Note that \citet{Ponti.2012} also reported a $\sigma^2_{\rm rms}$-$M_{\rm BH}$ relation for the timescale of 10 ks, but the dispersion is large, which is likely caused by the red noise leak, thus we do not try this timescale. Therefore, we derive a range of rms from different light curve segments of different lengths, which are listed in Table~\ref{tab-rms-mass}. For XMM-2, the rms is generally found to be 0.05-0.1 for all segments and all timescales, and the mean value of all the rms measurements is 0.071. The resultant range of black hole mass is $(0.8-2.5)\times10^{7}M_{\odot}$, and the mean value is $1.7\times10^{7}M_{\odot}$. This mass estimate and its range is fully consistent with the virial mass reported by \citet{Grupe.2010}. By comparison, the rms measured in XMM-1 is 0.232, which is a factor of 2-4 larger than XMM-2. The corresponding black hole mass is only $2.0\times10^{6}M_{\odot}$, which is almost one order of magnitude lower than the virial mass.

Therefore, this analysis also suggests that the rms in XMM-1 is over-estimated, which is probably caused by the red noise leak due to the short duration of this observation. This explanation should be checked by obtaining a much longer \xmm\ observation of \rxj0134\ in its X-ray high-flux state.

\subsection{Short-term UV Variability}
\label{sec-uvvar}
The strong short-term X-ray variability of \rxj0134\ also motivated us to check the optical/UV variability. During XMM-2, the OM camera was put in the {\tt Imaging+Fast} mode in the UVW1 filter, which has an effective wavelength of 2910 \AA, and so can record the UV flux simultaneously with the X-rays. Unfortunately, the {\tt Fast}-mode light curve is not valid due to an instrumental issue\footnote{During XMM-2, the source's PSF in OM drifted to the edge of the {\tt Fast}-mode field-of-view, which led to wrong flux measurements.}. Therefore, we can only produce a light curve by combining a series of exposures in the {\tt Imaging} mode, which has a much larger field-of-view, and so was not affected by the PSF drifting.

\begin{figure}
\centering
\includegraphics[trim=0.08in 0.4in 0.0in 0.1in, clip=1, scale=0.58]{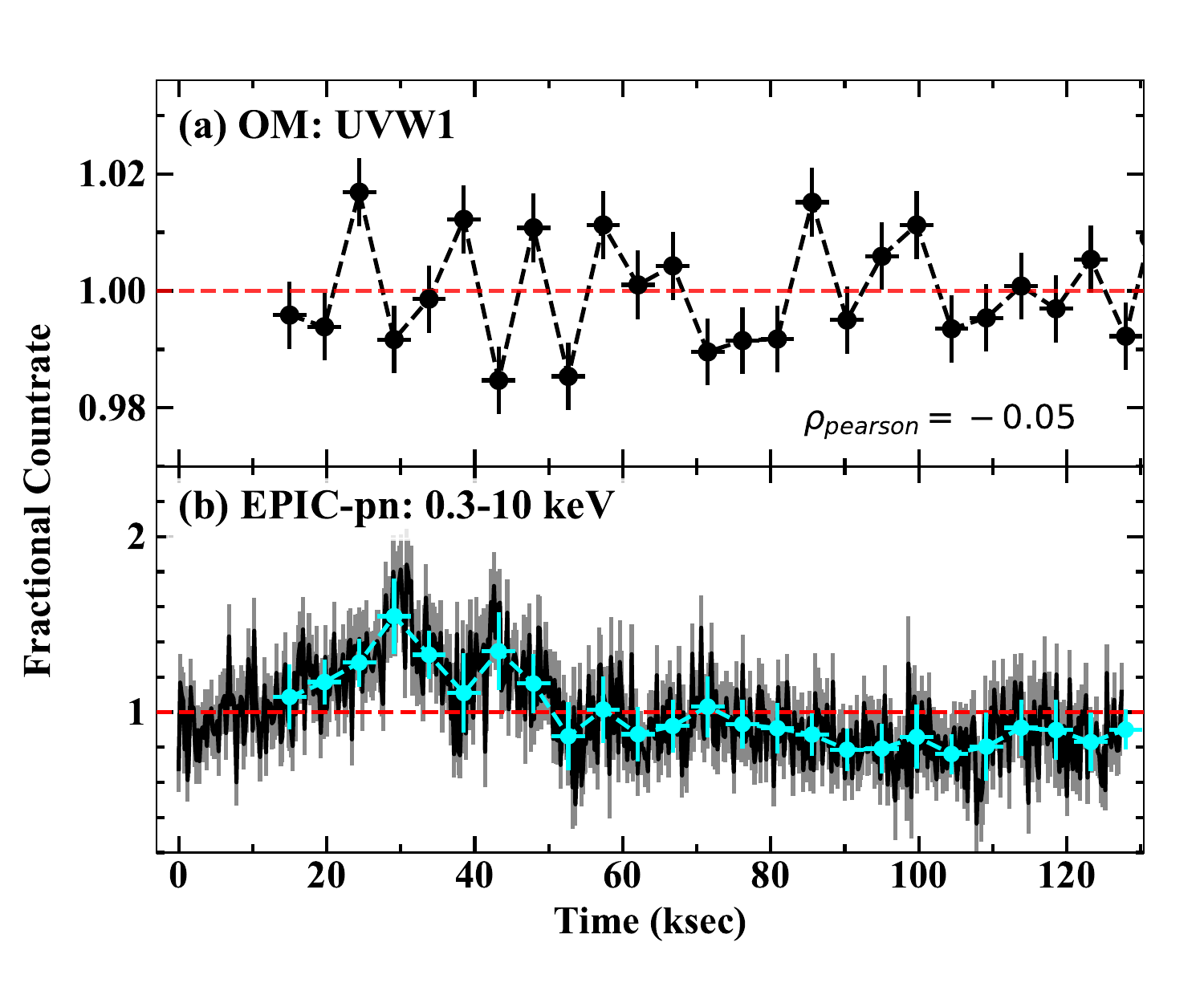} 
\caption{Comparison between the OM photometric light curve in UVW1 and the 0.3-10 keV light curve in EPIC-pn, which are observed simultaneously during the \xmm\ observation of \rxj0134\ in 2019. The apparent weak variability in the UVW1 light curve is mostly like due to the instrumental effect (see Section~\ref{sec-uvvar}). The X-ray light curve is also rebinned with the same binning time as OM, as shown by the cyan points. The Pearson's correlation coefficient is -0.05, indicating that there is no significant short-term UV/X-ray correlation.}
\label{fig-omvar}
\end{figure}

Figure~\ref{fig-omvar}a shows the short-term variability of \rxj0134\ in the UVW1 filter. The fractional rms amplitude of this UVW1 light curve is $0.7\pm0.1$ \%. However, OM data in the {\tt Imaging} mode in UVW1 also contain an instrumental systematic variation of $\lesssim1$\%, which is caused by the different frame-time in the {\tt Imaging}-mode exposures, which leads to differences in the coincidence loss correction\footnote{Private communication with the \xmm\ instrument team.}. Hence the apparent UV variability is probably not intrinsic to \rxj0134. Nevertheless, we plot the simultaneously 0.3-10 keV light curve in Figure~\ref{fig-omvar}b, and rebin it with the same time resolution as the OM light curve (cyan points). The fractional rms amplitude of this rebinned X-ray light curve is $12.4\pm4.5$ \%, which is much stronger than the UV variability. This is not surprising, as the UV flux is often dominated by the emission of accretion disc for super-Eddington AGN (e.g. \citealt{Jin.2017b}, \citealt{Jin.2021}), and the disc emission is expected to be much more stable than corona at short timescales. The potential contamination from host galaxy starlight is negligible, as confirmed by our detailed multi-wavelength spectral analysis (see Paper-II). Hence the stability of the UV emission should be intrinsic to the accretion process. The potential UV/X-ray correlation is checked by calculating the Pearson's correlation between the two light curves, which is found to be -0.05, thus no short-term UV/X-ray correlation is found.

\section{Conclusions}
\label{sec-conclusion}
We conducted a new multi-wavelength campaign from radio to optical/UV to X-rays on one of the most enigmatic super-Eddington NLS1s, namely \rxj0134, using both space and ground based telescopes. This campaign produced a large and rich dataset, and the detailed results will be published in a series of papers. In this first paper we reported the latest results from simultaneous \xmm\ and \nustar\ observations in 2019-12-19 as well as archival datasets. Our detailed X-ray spectral-timing study revealed many interesting properties of \rxj0134, which are summarized below.

\begin{itemize}
\item the \xmm\ and \nustar\ observations in 2019-12-19 caught \rxj0134\ in one of its lowest X-ray flux states.
\item the X-ray time-average spectra of \rxj0134\ exhibited dramatic variations in terms of both flux and spectral shape during the past 23 years, and no significant soft excess was detected in any of these spectra.
\item we compared the two \xmm\ observations in 2008 (XMM-1) and 2019 (XMM-2), and found that \rxj0134\ exhibited very different X-ray spectral-timing properties during its X-ray low and high-flux states, including much weaker X-ray variability in the low-flux state.
\item the time-average spectrum in XMM-2 showed a steep rise above 4 keV, indicating the presence of a separate hard X-ray component. Our time lag analysis also showed that this component lags behind the soft X-rays by $\sim$ 4 ks in the low frequency band of $[0.95,5]\times10^{-5}$ Hz. No other lags are found below 4 keV.
\item we considered a number of different models to fit the time-average, rms and lag spectra observed in XMM-1 and XMM-2. We find that the variability properties of \rxj0134\ are not consistent with some models such as ionized reflection, where one component contributes significant fluxes to both soft and hard X-rays.
\item the soft X-ray variability can be well modeled if there is a warm Comptonisation component below 1 keV, whose shape is similar to the one producing the soft excesses in normal X-ray {\it simple} super-Eddington NLS1s, but it is about an order of magnitude weaker. 
\item the hard X-ray variability is consistent with an origin of reflection or partial covering absorption from low ionisation material from within 100 $R_{\rm g}$. This can be interpreted as further evidence for the presence of a clumpy disc wind, which is expected in such a highly super-Eddington NLS1.
\end{itemize}

This study has deepened our understanding of the exceptional X-ray properties of \rxj0134. But in order to better constrain the presence of a soft excess component and to the origin of the X-ray mechanism of the spectral component above 4 keV, deeper observations are required during the high-flux state. In our following papers, we will present detailed multi-wavelength spectral-timing studies of \rxj0134\ (Papers-II and III).

\section*{Acknowledgements}
We thank the anonymous referee for providing valuable comments and suggestions which have improved the paper.
CJ thanks Peng Jiang, Hongyan Zhou, Weimin Yuan for helping with the coordination of observations.
We thank the teams of the {\it Swift} satellite and the Siding Spring Observatory for approving and conducting the target-of-opportunity observations, as well as helping with the data reduction. We thank the \xmm\ team for helping to investigate instrumental issues related to the OM data. \\
CJ acknowledges the National Natural Science Foundation of China through grant 11873054, and the support by the Strategic Pioneer Program on Space Science, Chinese Academy of Sciences through grant XDA15052100. We acknowledge the science research grants from the China Manned Space Project with NO.CMS-CSST-2021-B11. CD acknowledges the Science and Technology Facilities Council (STFC) through grant ST/T000244/1 for support. HL acknowledges the support by Chinese Postdoctoral Science Foundation (2021M693203), and the National Natural Science Foundation of China through grant 12103061.\\
This work is based on observations conducted by \xmm, an ESA
science mission with instruments and contributions directly funded by
ESA Member States and the USA (NASA). This work also makes use of data from the \nustar\ mission, a project led by the California Institute of Technology, managed by the Jet Propulsion Laboratory, and funded by the National Aeronautics and Space Administration. This research has made use of the XRT Data Analysis Software (XRTDAS) developed under the responsibility of the ASI Science Data Center (ASDC), Italy.

\section*{Data Availability}
The data underlying this article are publicly available from the High Energy Astrophysics Science Archive Research Center (HEASARC) at https://heasarc.gsfc.nasa.gov, the \xmm\ Science Archive (XSA) at https://www.cosmos.esa.int/web/xmm-newton/xsa, the Barbara A. Mikulski Archive for Space Telescopes (MAST) at https://mast.stsci.edu/portal/Mashup/Clients/Mast/Portal.html, the Sloan Digital Sky Survey (SDSS) at http://skyserver.sdss.org/dr12/en/home.aspx. The SSO optical spectrum in this article will be shared on reasonable request to the corresponding author.











\bsp	
\label{lastpage}
\end{document}